\documentclass[]{spie}  %>>> use for US letter paper
\usepackage{amsmath,amsfonts,amssymb}
\usepackage{graphicx}
\usepackage[colorlinks=true, allcolors=blue]{hyperref}
\usepackage{siunitx}
\usepackage[]{xspace}
\usepackage{enumitem}
\usepackage{perpage} %the perpage package
\MakePerPage{footnote}
\usepackage [english]{babel}
\usepackage [autostyle, english = american]{csquotes}
\def\bi{\begin{itemize} \itemsep1pt \parskip2pt \parsep0pt}
\def\ei{\end{itemize}}

\def\degr{\hbox{$^\circ$}}
\def\arcsec{\hbox{$^{\prime\prime}$}}
\newcommand{\micron}{$\SI{}{\, \micro\meter}$\xspace}

\newcommand{\jwst}{{\tt \href{https://jwst-pipeline.readthedocs.io/en/latest/}{jwst}}\xspace}

\newcommand{\nircam}{{\tt \href{https://jwst.stsci.edu/instrumentation/nircam}{NIRCam}}\xspace}
\newcommand{\miri}{{\tt \href{https://jwst.stsci.edu/instrumentation/miri}{MIRI}}\xspace}

\newcommand{\pynrc}{{\tt \href{https://pynrc.readthedocs.io/en/latest/}{pyNRC}}\xspace}
\newcommand{\nirccos}{{\tt \href{https://github.com/kammerje/NIRCCoS}{NIRCCoS}}\xspace}

\newcommand{\pancake}{{\tt \href{https://github.com/spacetelescope/pandeia-coronagraphy}{PanCAKE}}\xspace}

\newcommand{\spaceklip}{{\tt \href{https://github.com/kammerje/spaceKLIP}{spaceKLIP}}\xspace}
\newcommand{\pyklip}{{\tt \href{https://pyklip.readthedocs.io}{pyKLIP}}\xspace}
\newcommand{\coronsup}{{\tt \href{https://www.stsci.edu/jwst/science-execution/program-information.html?id=1441}{PID 1441}}\xspace}
\newcommand{\coronta}{{\tt \href{https://www.stsci.edu/jwst/science-execution/program-information.html?id=1075}{PID 1075}}\xspace}
\newcommand{\coronlmc}{{\tt \href{https://www.stsci.edu/jwst/science-execution/program-information.html?id=1070}{PID 1070}}\xspace}

\newcommand{\mast}{{\tt \href{https://archive.stsci.edu}{MAST}}\xspace}

\newcommand{\hci}{{\tt \href{https://jwst-docs.stsci.edu/display/JPP/JWST+High-Contrast+Imaging?q=High}{HCI}}\xspace}% 
\newcommand{\ers}{{\tt \href{https://jwst.stsci.edu/observing-programs/approved-ers-programs}{ERS}}\xspace}%
\newcommand{\gto}{{\tt \href{https://jwst.stsci.edu/observing-programs/approved-gto-programs}{GTO}}\xspace}%
\newcommand{\webbpsf}{{\tt \href{https://webbpsf.readthedocs.io/en/latest/}{WebbPSF}}\xspace}%
\newcommand{\webbpsfext}{{\tt \href{https://github.com/JarronL/webbpsf_ext}{WebbPSF\_{ext}}}\xspace}%
\newcommand{\mirage}{{\tt \href{https://github.com/spacetelescope/mirage}{MIRaGe}}\xspace}%

 \newcommand{\jwstdistortion}{{\tt \href{https://github.com/tonysohn/jwst_fpa}{jwst\_distortion.py}}\xspace}
 \newcommand{\pysiaf}{{\tt \href{https://pysiaf.readthedocs.io/en/latest/}{pySIAF}}\xspace}

\def\begini{\begin{itemize}[itemsep=0.5pt,topsep=0.5pt]}
\def\endi{\end{itemize}}

\title{JWST/NIRCam Coronagraphy:\\ Commissioning and First On-Sky Results}

\author[a]{Julien H. Girard}
\author[b]{Jarron Leisenring}
\author[a]{Jens Kammerer}
\author[a]{Mario Gennaro}
\author[b]{Marcia Rieke}
\author[a]{John Stansberry}
\author[a]{Armin Rest}
\author[b]{Eiichi Egami}
\author[a]{Ben Sunnquist}
\author[a]{Martha Boyer}
\author[a]{Alicia Canipe}
\author[a]{Matteo Correnti}
\author[a]{Bryan Hilbert}
\author[a]{Marshall D. Perrin}
\author[a]{Laurent Pueyo}
\author[a]{Remi Soummer}
\author[a]{Marsha Allen}
\author[a]{Howard Bushouse}
\author[a]{Jonathan Aguilar}
\author[a]{Brian Brooks}
\author[a]{Dan Coe}
\author[a]{Audrey DiFelice}
\author[a]{David Golimowski}
\author[a]{George Hartig}
\author[a]{Dean C. Hines}
\author[a]{Anton Koekemoer}
\author[a]{Bryony Nickson}
\author[a]{Nikolay Nikolov}
\author[a]{Vera Kozhurina-Platais}
\author[a]{Nor Pirzkal}
\author[a]{Massimo Robberto}
\author[a]{Anand Sivaramakrishnan}
\author[a]{Sangmo Tony Sohn}
\author[a]{Randal Telfer}
\author[a]{Chi Rai Wu}
\author[b]{Thomas Beatty}
\author[b]{Michael Florian}
\author[b]{Kevin Hainline}
\author[b]{Doug Kelly}
\author[b]{Karl Misselt}
\author[b]{Everett Schlawin}
\author[b]{Fengwu Sun}
\author[b]{Christina Williams}
\author[b]{Christopher Willmer}
\author[c]{Christopher Stark}
\author[d]{Marie Ygouf}
\author[d,f]{Charles Beichman}
\author[e]{Aarynn Carter}
\author[g]{Thomas P.  Greene}
\author[g]{Thomas Roellig}
\author[d]{John Krist}
\author[h]{J\'ea Adams Redai}
\author[i,j]{Jason Wang}
\author[k]{Charles R. Clark}
\author[l]{Dan Lewis}
\author[l]{Malcolm Ferry}
\affil[a]{\small Space Telescope Science Institute (STScI), 3700 San Martin Dr, Baltimore MD, 21218, USA}
\affil[b]{\small Steward Observatory, Univ. of Arizona, Tucson, 933 N Cherry Ave., Tucson, AZ 85721, USA}
\affil[c]{\small NASA Goddard Space Flight Center, Greenbelt, MD 20771, USA}
\affil[d]{\small Jet Propulsion Laboratory (JPL), California Institute of Technology, Pasadena, CA 91109, USA}
\affil[e]{\small Department of Astronomy \& Astrophysics, University of California, Santa Cruz, CA 95064, USA}
\affil[f]{Caltech//IPAC/NASA Exoplanet Science Institute (NExScI), Pasadena, CA 91125, USA}
\affil[g]{\small NASA Ames Research Center, Mountain View, CA 94035, USA}
\affil[h]{\small CFA, Harvard \& Smithsonian, 60 Garden Street, Cambridge, MA 02138, USA}
\affil[i]{\small Department of Astronomy, California Institute of Technology, Pasadena, CA 91125, USA}
\affil[j]{\small CIERA and Department of Physics \& Astronomy, Northwestern University, Evanston, IL 60208, USA}
\affil[k]{\small AK Aerospace Technology Corporation, 6301 Ivy Ln \#700, Greenbelt, MD 20770, USA}
\affil[l]{\small Lockheed Martin Advanced Technology Center (ATC), 3251 Hanover St, Palo Alto, CA 94304, USA}
\authorinfo{Further author information: send correspondence to jgirard@stsci.edu}

\begin{document} 
\maketitle

\begin{abstract}
%\todo{NEED TO BE UPDATED}
In a cold and stable space environment, the James Webb Space Telescope (JWST or "Webb") reaches unprecedented sensitivities at wavelengths beyond 2 microns, serving most fields of astrophysics. It also extends the parameter space of high-contrast imaging into the near and mid-infrared. Launched in late 2021, JWST underwent a six month commissioning period. In this contribution we focus on the NIRCam Coronagraphy mode which was declared "science ready" on July 10 2022, the last of the 17 JWST observing modes. Essentially, this mode enables the detection of fainter/redder/colder (less massive for a given age) self-luminous exoplanets as well as other faint astrophysical signal in the vicinity of any bright object (stars or galaxies). Here we describe some of the steps and hurdles the commissioning team went through to achieve excellent performance. Specifically, we focus on the Coronagraphic Suppression Verification activity. We were able to produce firm detections at 3.35$\mu$m of the white dwarf companion HD 114174 B which is at a separation of $\simeq$ 0.5\arcsec and a contrast of $\simeq$ 10 magnitudes ($10^{4}$ fainter than the K$\sim$5.3 mag host star). We compare these first on-sky images with our latest, most informed and realistic end-to-end simulations through the same pipeline. Additionally we provide information on how we succeeded with the target acquisition with all five NIRCam focal plane masks and their four corresponding wedged Lyot stops.

\end{abstract}

% Include a list of keywords after the abstract 
\keywords{High Contrast Imaging, Infrared Astronomy, Coronagraphy, James Webb Space Telescope (JWST), Commissioning, NIRCam, Exoplanets, High Angular Resolution}

\section{NIRCam Coronagraphs Are Science Ready}
\label{sec:intro}  % \label{} allows reference to this section

\begin{figure}[h!]
\begin{center}
\includegraphics[width=0.72\textwidth, angle=0]{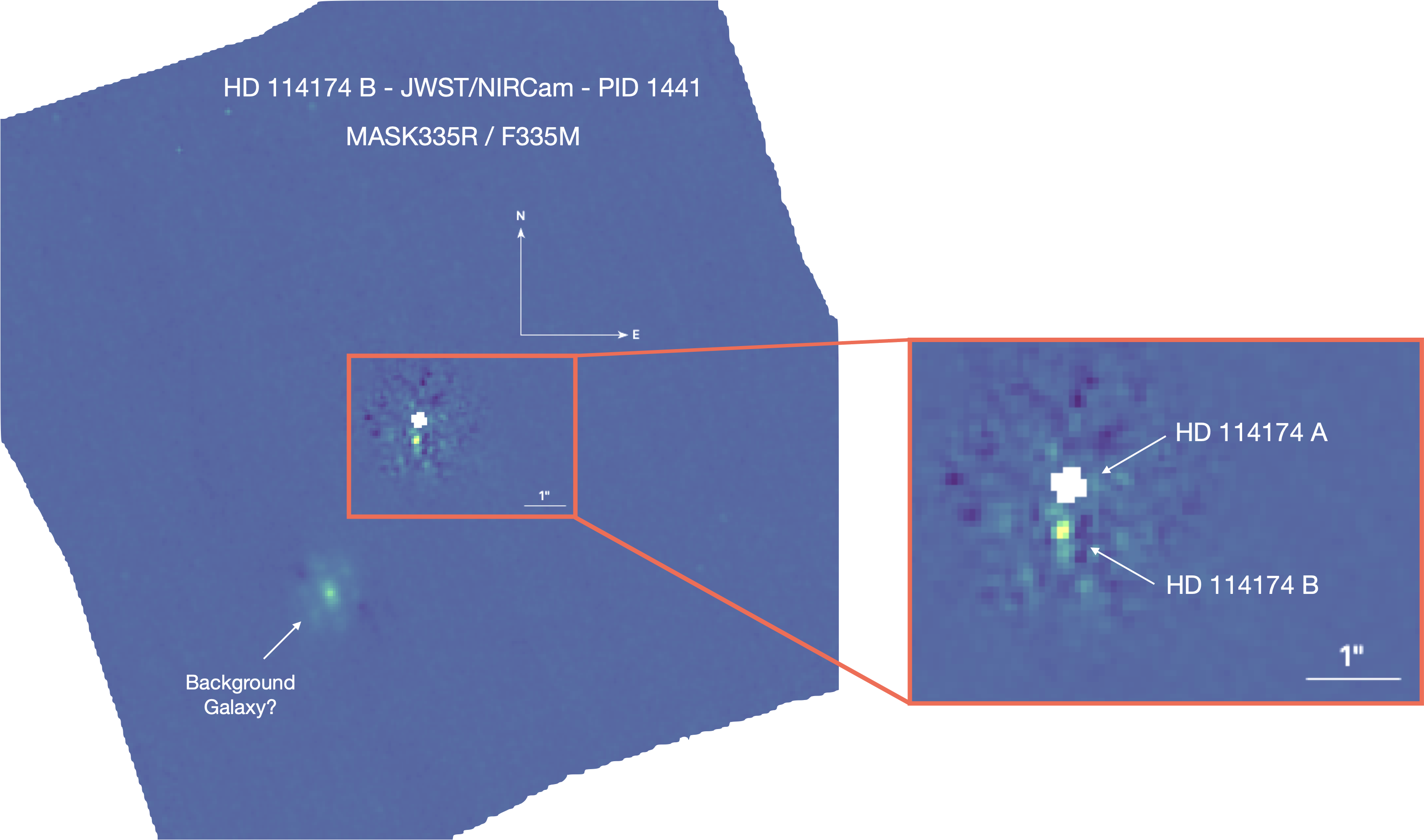}
\vspace{0.2cm}
\caption{Demonstration of the \nircam Coronagraphy potential with the clean imaging of the WD companion HD 114174 B, 10,000 times fainter than HD 114174 A (subtracted out by means of reference stars). This was taken with the round MASK335R through the F335M filter by the \nircam team during commissioning on July 5 2022, \coronsup (PI J. Girard).}
\label{fig:on-sky-335r}
\end{center}
\end{figure}

Before diving into details about how we prepared and carried out the commissioning of the \nircam Coronagraphy mode\footnote{Landing page "NIRCam Coronagraphic Imaging" on the JWST User Documentation platform/wiki (JDox)\cite{jdox_general}:  {\small \tt \href{https://jwst-docs.stsci.edu/jwst-near-infrared-camera/nircam-observing-modes/nircam-coronagraphic-imaging}{jwst-docs.stsci.edu/jwst-near-infrared-camera/nircam-observing-modes/nircam-coronagraphic-imaging}}}, figure~\ref{fig:on-sky-335r} displays an on-sky image and detection of the white dwarf (WD) HD 114174 B, companion to the $\sim$4 billion year old main sequence star HD 114174A (spectral type G5IV-V). Together they form a \enquote{Sirius like system} \cite{gratton2021}. This WD was first imaged about 10 years ago with the rise of adaptive optics (AO) \cite{crepp2013} and it is used a spectrophotometric calibrator for extreme AO instruments like SPHERE \cite{beuzit2019}. With a contrast of $\simeq$ 10 magnitudes ($10^{4}$) and a separation of 0.5\arcsec (the companion has gotten closer in recent years), it is also a \enquote{perfect} object to use to demonstrate the high contrast imaging (HCI) capability of a new instrument and/or telescope. Figure~\ref{fig:on-sky-335r} shows that the detection has a high signal to noise ratio (SNR) as the recovered WD signal harbors the six secondary spots typical of the round mask point spread function (PSF) with an aperture is defined by the Lyot stop as seen in figure~\ref{fig:ta}.

\nircam coronagraphy will be used primarily to directly image and characterize young, self-luminous giant exoplanets\cite{bowler2016_review} and their circumstellar environment (protoplanetary disks, debris disks, jets). At short separations (e.g. $\leqslant$0.3\arcsec at 2\micron, $\leqslant$0.5\arcsec at 3.5\micron) \nircam will not outperform ground based extreme AO-fed instruments on 8 to 10-meter telescopes. But at larger separations and wavelengths the stability and sensitivity will outperform even future extremely large telescopes (25 to 39-meter class). All these world class observatories will be very complementary\cite{girard2020_muse}. There are already a lot of synergies between ground and space, ALMA (submillimetric) and Hubble and now Webb. Coronagraphy on-board JWST (\nircam and \miri)\cite{girard2018spie} - because of its stability in space - will also be possible for rather faint young stellar objects (YSO) and extragalactic targets (e.g. active galactic nuclei, etc.).

\subsection{NIRCam Coronagraphy: How Does This Mode Work?}
\label{sec:nrc-coron}  % \label{} allows reference to this section

Due to its complexity and dependency on other observatory functionalities, \nircam Coronagraphy was one of the last modes to be commissioned. It is highly sensitive to and less forgiving than other modes to Target Acquisition (TA) error, guiding accuracy, wavefront error (WFE), focus and distortion correction. The distortion correction is particularly difficult to achieve and must be performed independently of the Imaging Mode solution. In addition, with five coronagraphs to commission, it is a lengthy process.

\begin{figure}[h!]
\begin{center}
\includegraphics[width=0.95\textwidth, angle=0]{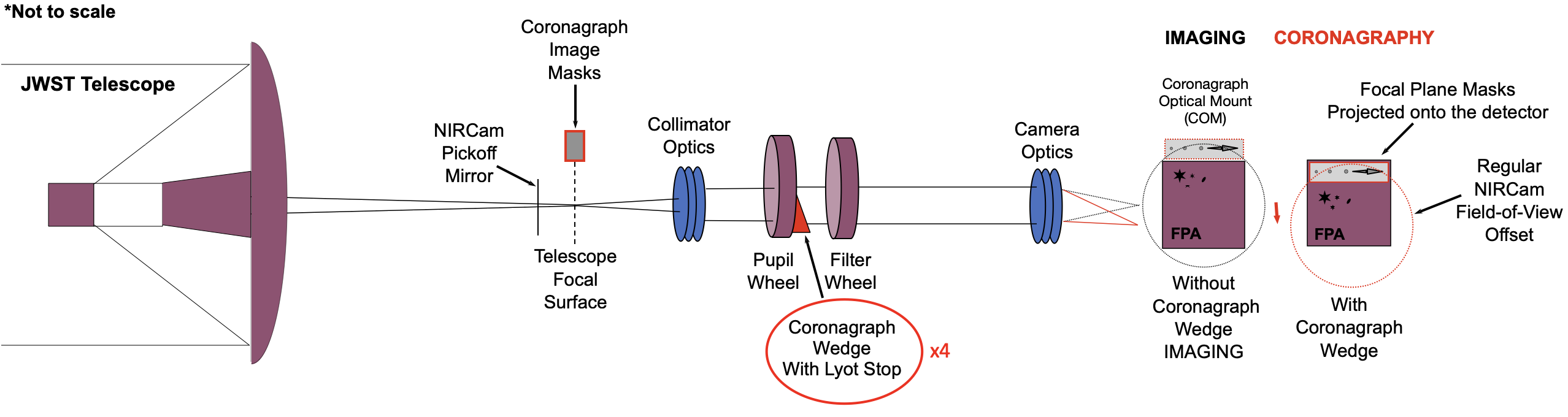}
\caption{Schematic view of the optical principle of the NIRCam coronagraphs. When the wedge (together with the Lyot stop in the pupil wheel) is in the beam, the whole field of view (FoV) shifts down. The Coronagraphic masks (introduced in the beam upstream)  which are out of the nominal Imaging mode FoV are now imaged onto the focal plane array (FPA).} 
\label{fig:principle}
\end{center}
\end{figure}

The \nircam coronagraphs  were designed\cite{krist2007_spie_disk, krist2009, krist2010_spie_jwst_occulters} to work in the face of significant diffraction from the Optical Telescope Element (OTE) segments and in the face of possible pupil shear.  The Lyot stops have undersized holes to solve these problems. Additionally, the coronagraph was designed to work outside of the regular field of view used for wavefront sensing and surveys. This requirement was met by making the four Lyot stops wedged to deflect the field of view as explained in figure~\ref{fig:principle}. This is important to note that there are four such wedge$+$Lyot stop elements: SW RND, SW BAR, LW RND (both for MASK335R and MASK430R), LW BAR. Each of them introduces a slightly different offset and a different distortion and pupil \enquote{wander} (each field point has a slightly different pupil alignment between \nircam and the OTE) and therefore  we needed to compromise pupil element positioning between the different round masks (specifically M335R and M430R which share the same Lyot stop). 
This has implications both to achieve precise target acquisition and to provide astrometrically calibrated data to the community.

\subsection{Science Readiness Criteria}
\label{sec:sr}  % \label{} allows reference to this section

There were no contractual requirements for the \nircam Coronagraphy mode. Nevetherless, the relevant metrics agreed between the PI and commissioning scientists at STScI and NASA Goddard were:
\begin{enumerate}
\item Contrast: 5-$\sigma$ contrast of $10^4$ at 1\arcsec with the F335M filter and the MASK335R (most versatile round mask for LW) with reference star subtraction (Reference Differential Imaging: RDI).
\item Target Acquisition: better than 0.5 pixels (1-$\sigma$) for any coronagraphic mask ($\leqslant$15 mas for SW, $\leqslant$30 mas for LW).
\end{enumerate}

If the telescope managed to deliver a stable and diffraction limited images at $\leqslant$ 2 \micron and we managed to perform target acquisition to within a pixel or so, we knew from recent simulations that the first criterion would be easily met. The second criterion would guarantee even better, expected performance\cite{perrin2018, carter2021, carter2021spie, hinkley2022simulations}. 

% NIRCam meets the first of these easily but is borderline with the second 
%Performance is what matters most for NIRCam coronagraphy: Contrast

\section{Preparation: simulation and astrometric framework}
\label{sec:prep}  % \label{} allows reference to this section

During the few years preceding the JWST launch, instrument teams have been rehearsing and getting ready for the commissioning by exercising proposal preparation tools, data reduction \jwst pipeline\footnote{JWST Data Analysis With the JWebbinars: {\small \tt \href{https://www.stsci.edu/jwst/science-execution/jwebbinars}{stsci.edu/jwst/science-execution/jwebbinars}}}\cite{gordon2022_JWSTCAL} and analysis scripts on simulated data. \nircam Coronagraphy is no exception and Girard et al. 2018\cite{girard2018spie} described the \enquote{end to end} prototype developed then. Of course, when real data \enquote{come down}, things can differ a bit and the team took advantage of having \nircam be operated from the start of OTE commissioning (as a camera for all the telescope / wavefront sensing activities) to solve a number of issues and characterize all 10 detectors (or Sensor Chip Assembly: SCAs).

\subsection{pyNRC}
\label{sec:pynrc}  % \label{} allows reference to this section

\pynrc is \enquote{a set of Python-based tools for planning observations with JWST NIRCam. It includes an Exposure Time Calculator (ETC), a simple image slope simulator, and an enhanced data simulator compatible with the JWST pipeline. This package works for a variety of NIRCam observing modes including direct imaging, coronagraphic imaging, slitless grism spectroscopy, and weak lens imaging. All PSFs are generated via \webbpsf and \webbpsfext (extensions) to reproduce realistic JWST images and spectra}. 

\noindent For \nircam Coronagraphy \pynrc has been instrumental. While early end-to-end data prototypes\cite{girard2018spie} were made using a set of packages (\pancake, \webbpsf, \mirage). Approaching commissioning, the team relied almost entirely on \pynrc which, with \webbpsfext,  integrates everything (interface to APT files and catalogs including Gaia, ramp simulator with noise sources and cosmic rays, \pysiaf apertures, generation of \jwst pipeline compliant products). 

\begin{figure}[h!]
\begin{center}
\includegraphics[width=0.9\textwidth, angle=0]{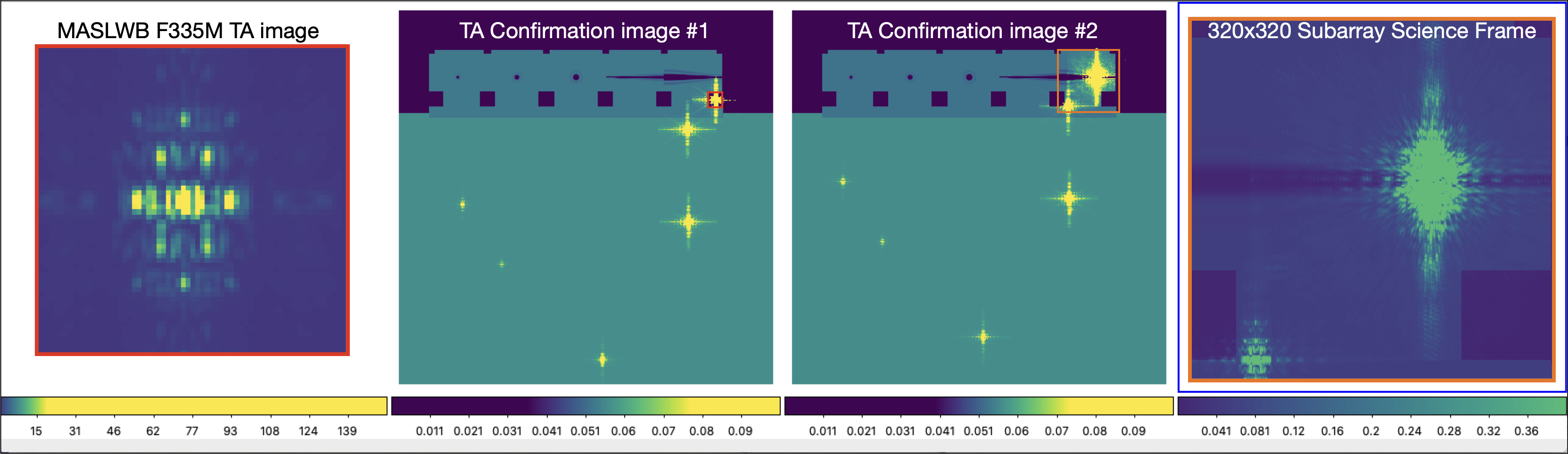}
\vspace{0.2cm}
\caption{Example of simulations of a program making use of the LW bar mask, noiseless \enquote{slope} images are very useful (masks and ND filters are clearly visible) to make sure the we are simulating the correct, expected scene. Left: unnocculted TA image. Center (two images): TA astrometric confirmation image \#1 after the TA centro\"iding and before the SAM to place the star behind the mask. TA astrometric confirmation image \#2 once the SAM has been performed and the star of interest is located where it should be, behind the mask. Right: science frame in subarray mode. \pynrc fetched field stars from Gaia so one can know in advance whether absolute astrometry can be performed and which stars can be used to do relative astrometry and determine the true position of the star behind the mask using the movement of other field stars.}
\label{fig:pynrc}
\end{center}
\end{figure}

Figure~\ref{fig:pynrc} shows an examples of noiseless (\texttt{slope}), yet useful \pynrc simulations. All of these very generated directly from APT exported files, something that the \nirccos wrapper\cite{kammerer2022spie} also does. \pynrc will also generate noisy \texttt{uncal} files with all the DMS compliant headers for the \jwst pipeline stages to run. \pynrc can also generate a scenario of observations, introduce a sensible wavefront drift between science target and reference star(s) as well as compute predicted contrasts.

\begin{figure}[h!]
\begin{center}
\includegraphics[width=0.7\textwidth, angle=0]{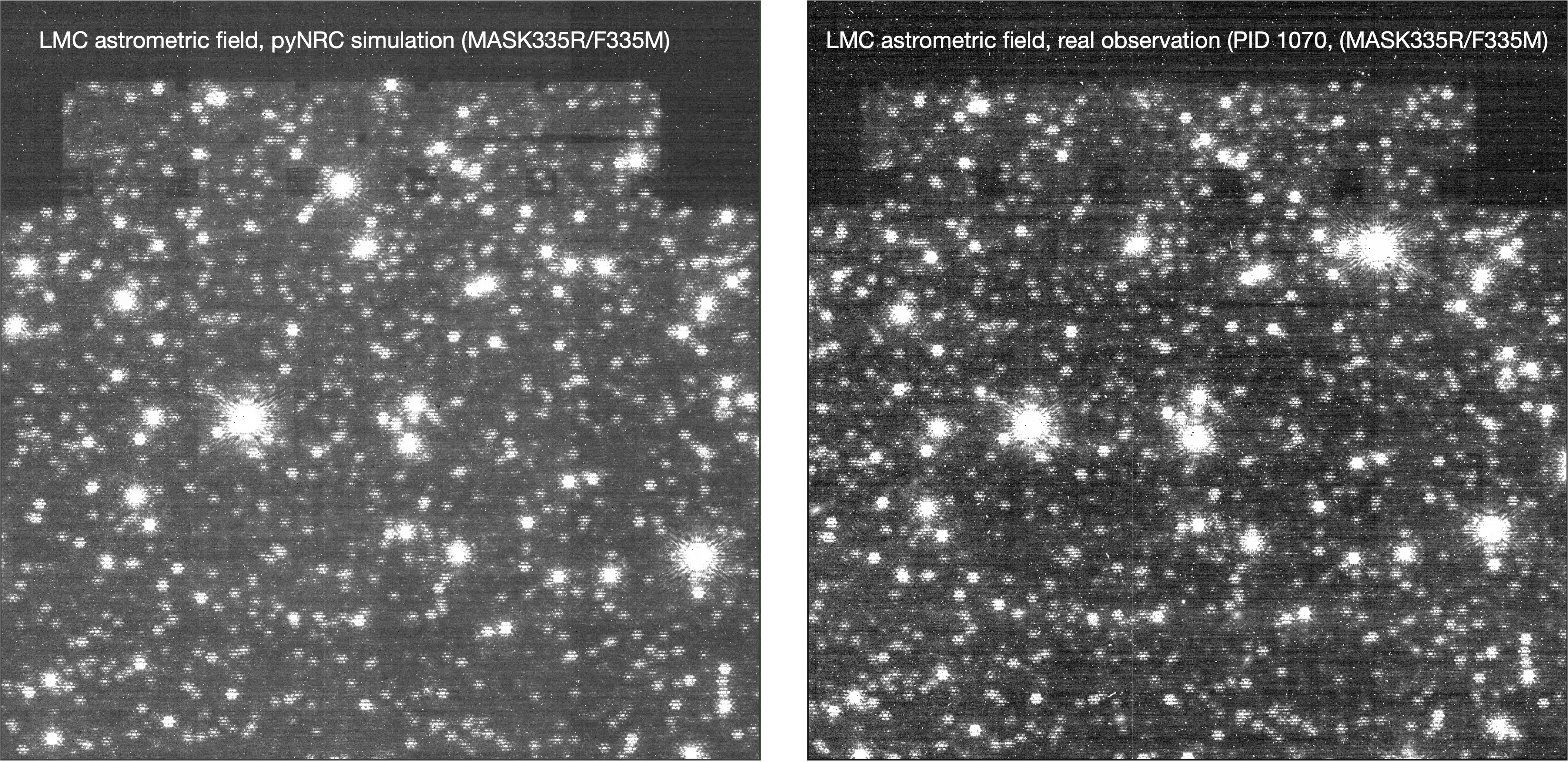}
\vspace{0.2cm}
\caption{Left: \pynrc full frame A5 MASK335R/F335M simulation of the LMC astrometric field. Right: real, on-sky image obtained as part of our coronagraphic astrometric program (\coronlmc same as in figure~\ref{fig:astrom1}) and reduced by the stage 1 and 2 \jwst pipeline. There is only a difference in position angle of a few degrees.}
\label{fig:pynrc-lmc}
\end{center}
\end{figure}

\subsection{WebbPSF/WebbPSF{\_ext}}
\label{sec:webbpsf}  % \label{} allows reference to this section

During the Science Instruments (SI) commissioning, the OTE team had more and more experience in performing routine maintenance measurements of the wavefront / optical path differences (OPD) maps of the telescope using the fine phasing technique provided by the \nircam weak lens mode (providing defocused images for phase retrieval). \webbpsf was modified to allow the use of contemporary OPDs measured on orbit\footnote{JWST Using OPDs Measured On Orbit: {\small \tt \href{https://webbpsf.readthedocs.io/en/latest/jwst_measured_opds.html}{webbpsf.readthedocs.io/en/latest/jwst\_measured\_opds.html}}} within days of (before or after) any given program or activities. This proved to be extremely useful for our \nircam Coronagraphy commissioning subteam as we heavily relied on high fidelity simulations to assess the TA accuracy and generally succeed with it.

\noindent Calling \webbpsf \enquote{on the fly} can be computationally intensive and \webbpsfext \enquote{provides some enhancements to the \webbpsf package for PSF creation. This follows the \pynrc implementation for storing and retrieving JWST PSFs. In particular, this module generates and saves polynomial coefficients to quickly create unique instrument PSFs as a function of wavelength, focal plane position, wavefront error drift from thermal distortions. More specifically, \webbpsfext uses \webbpsf to generate a series of monochromatic PSF simulations, then produces polynomial fits to each pixel. Storing the coefficients rather than a library of PSFs allows for quick creation (via matrix multiplication) of PSF images for an arbitrary number of wavelengths (subject to hardware memory limitations, of course). The applications range from quickly creating PSFs for many different stellar types over wide bandpasses to generating a large number of monochromatic PSFs for spectral dispersion}.

\subsection{SIAF Infrastructure}
\label{sec:siaf}  % \label{} allows reference to this section

\begin{figure}[h!]
\begin{center}
\includegraphics[width=0.55\textwidth, angle=0]{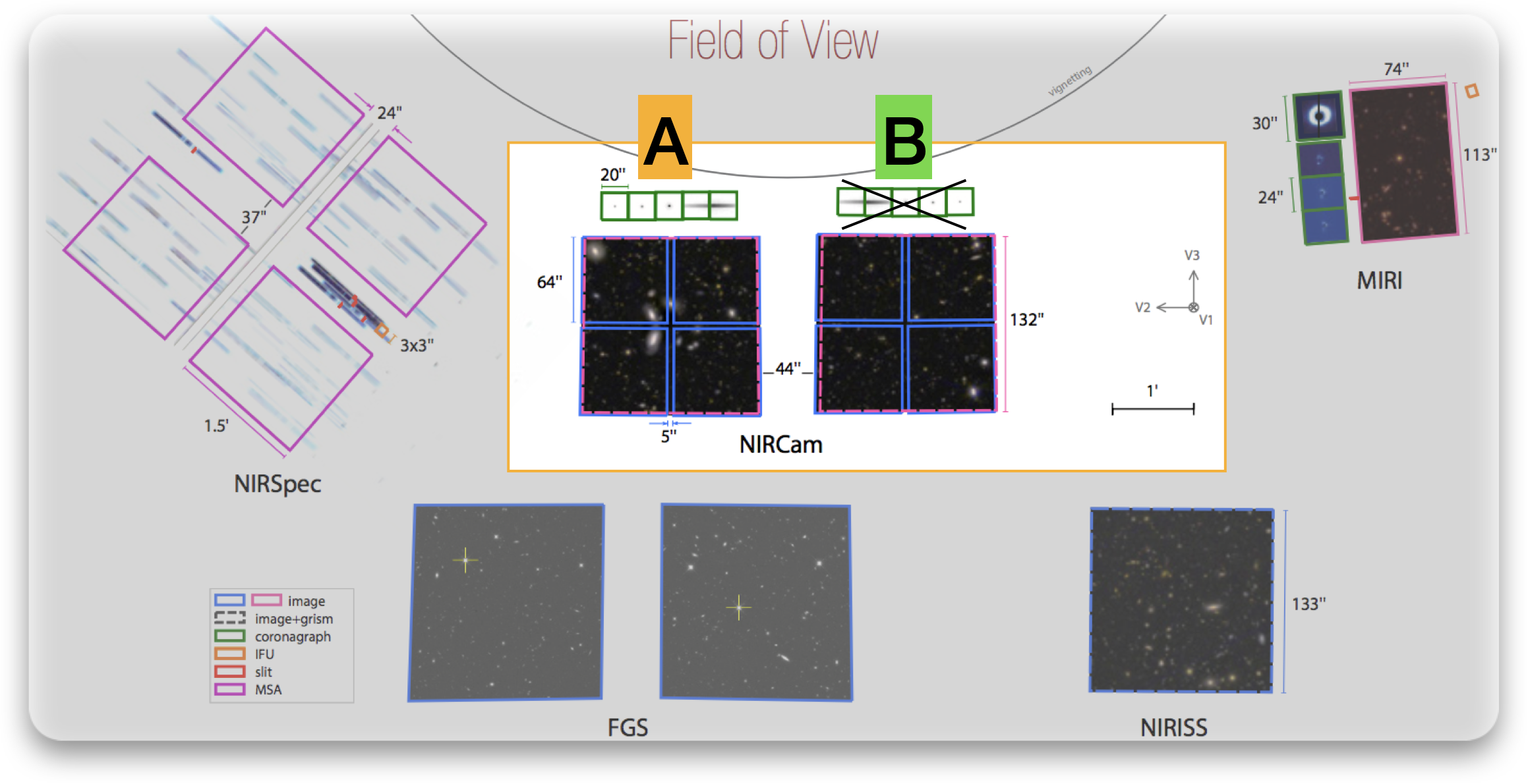}
\includegraphics[width=0.3\textwidth, angle=0]{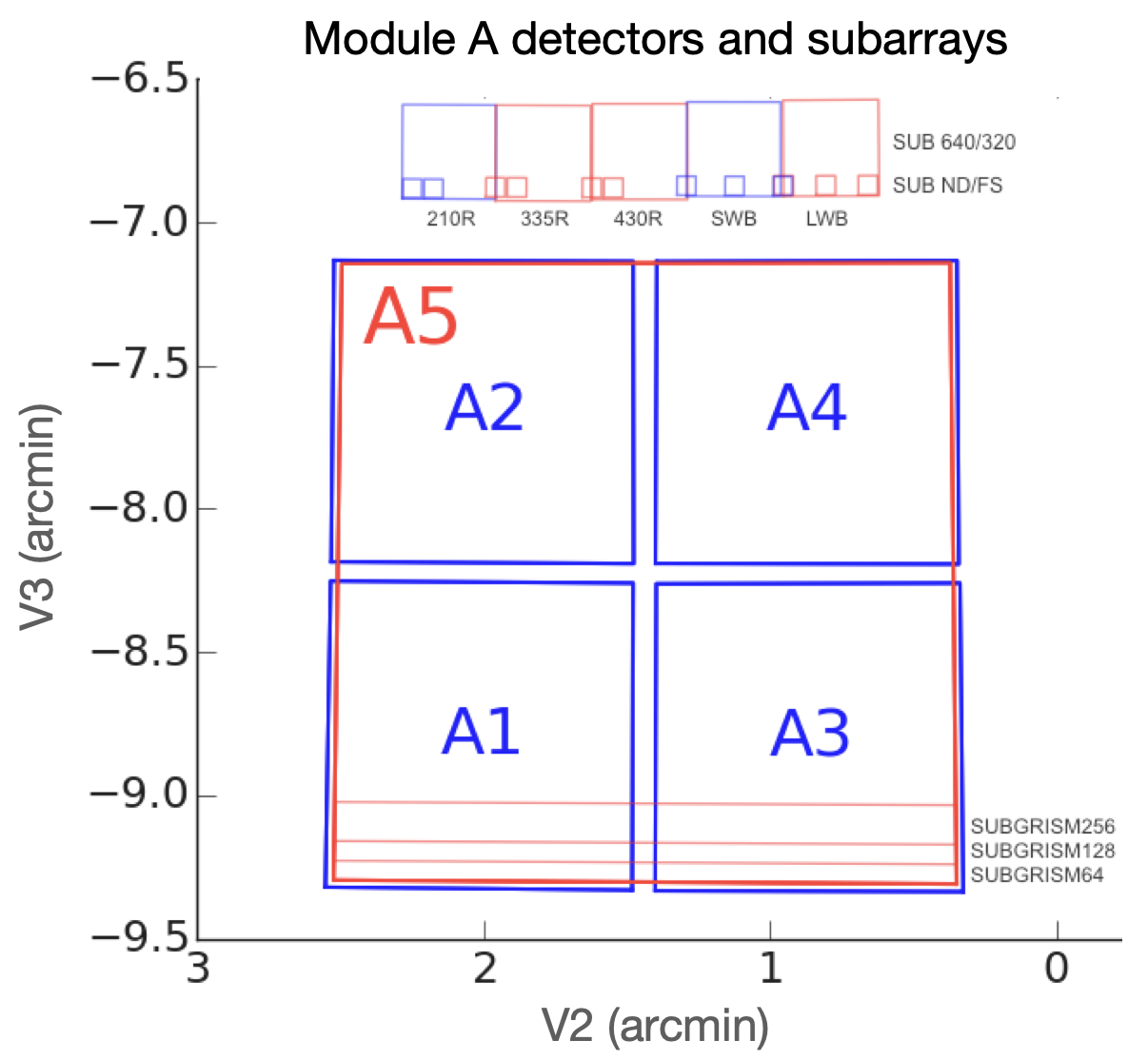}
\vspace{0.15cm}
\caption{Left: science instruments\cite{rieke2005_NIRCam, rieke2015_MIRI, jakobsen2022_NIRSpec, doyon2012_FGS} fields of view in the JWST focal plane, only \nircam Module A is enabled for Coronagraphy. Right: The 5 SCAs of Module A (A5 for LW and A2-A2-A3-A4 for SW). The SIAF framework allows to assign the correct WCS to each pixel in the focal plane.}
\label{fig:siaf}
\end{center}
\end{figure}

Science Instrument Aperture File (SIAF) is a reference file used in operations that contains the official information on all apertures (e.g., \nircam Apertures) and internal instrument coordinates. For instance:
\bi
\item ($V2_\textrm{Ref}$, $V3_\textrm{Ref}$) is the reference position in ($V2$, $V3$) coordinates (arcsec); some of these entries are used to define telescope pointings.
\item $V3_\textrm{Idl Y Angle}$ is the rotation (in degrees, counterclockwise) of the aperture's ideal Coordinate System Y-axis relative to $V3$.
\item The ideal coordinate system is a distortion-removed frame used for dithers and other pointing offsets. These coordinates correspond to a functional transform of the pixel coordinates in the science frame. The orientation and parity of the ideal coordinate system are equal to the pixel coordinates.
\item ($V2_1$, $V2_2$, $V2_3$, $V2_4$), ($V3_1$, $V3_2$, $V3_3$, $V3_4$) are the vertices in the ($V2$, $V3$) coordinates (arcsec) of the quadrilateral defined by each aperture.
\ei

Each of our deliveries to update the SIAF (e.g. mask positions, offsets with respect to FGS) had to be carefully crafted and verified. Simple formatting mistakes or a mis-transformation  of coordinates can result in a completely wrong TA offset (SAM) calculation and a loss of precious days of commissioning and telescope time. 

\subsection{NIRCam Commissioning Activities}
\label{sec:nrc-com}  % \label{} allows reference to this section

\begin{figure}[h!]
\begin{center}
%\includegraphics[width=0.45\textwidth, angle=0]{FiguresSPIE/pseudo_SGD_contrast_Obs15_MASK210R.png}
%\hfill
\includegraphics[width=0.8\textwidth, angle=0]{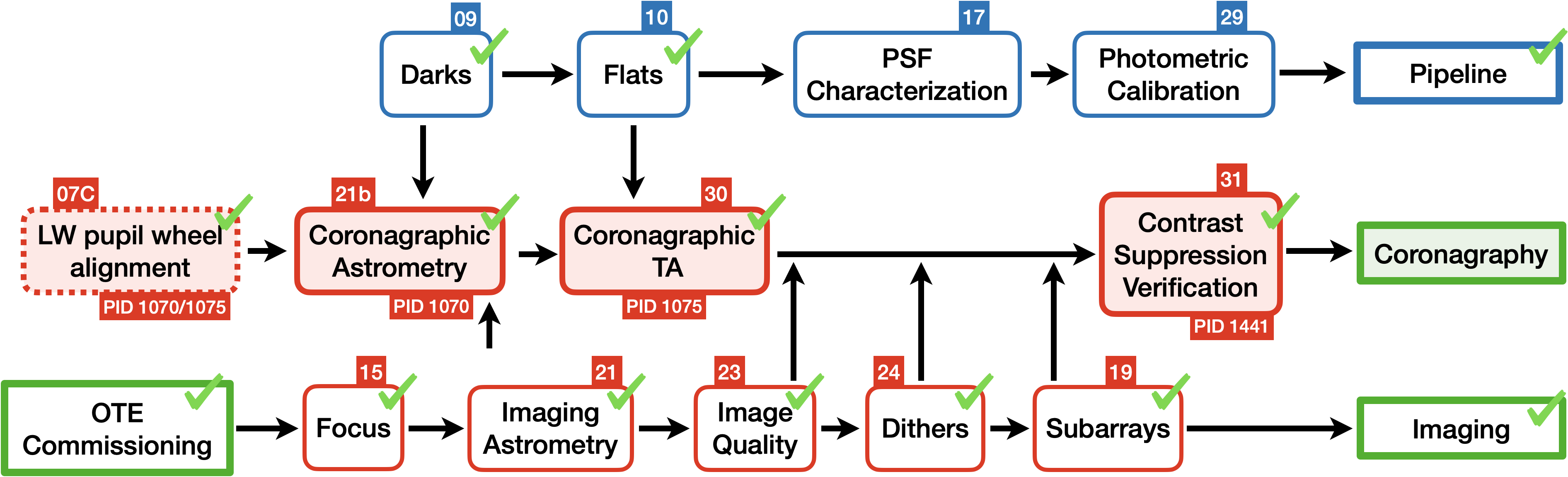}
\vspace{0.25cm}
\caption{Main commissioning activities for NIRCam Coronagraphy with their links to all other activities which need a successful execution for this mode.}
\label{fig:nrc-cars}
\end{center}
\end{figure}

\nircam was used to acquire the first photons of JWST and to align and focus all 18 telescope segments. \nircam took all the \enquote{selfies} of the primary mirror with its dedicated pupil imaging lens (PIL) in the short wavelength (SW) channel. \nircam was thus the first science instrument (SI) to be used from the end of January 2022, a month after launch. As coronagraphy is one of the most complex modes of the instrument and has dependencies on many other activities, it was expected to be one of the last modes to be declared \enquote{science ready}. Figure~\ref{fig:nrc-cars} describes the commissioning activities related or needed to check out the \nircam Coronagraphy mode.

\subsection{The Coronagraphic Suppression Verification Program: PID 1441}
\label{sec:nrc-31} 

This program was largely inspired by simulation work in Perrin et al. 2018 showing that deeper contrasts can be achieved using more than one PSF reference star\cite{perrin2018}. In 2018-2019 we built a case for such a program as no programs had been approved to qualify \nircam Coronagraphy performance other than the TA one with a very limited possibility to explore the contrast. We therefore decided to design a program whose main goal is to demonstrate that we can perform effective coronagraphic, star-light suppression with JWST/NIRCam and achieve the performance expected and simulated thus far: contrasts, inner-working angles (IWA) and detection limits (at various angular distances, especially in the speckle limited regime). Our hopes were to achieve a \enquote{desired / expected performance} significantly better than the minimum contrast \enquote{science readiness} requirement, reach 5$\times$10$^{-6}$ below 1\arcsec. A second important objective is to be able to provide clear guidelines to observers through a post-commissioning update of our high-contrast documentation suite (JDox, etc.) and tuning of the Exoposure Time Calculator (ETC) and other tools. For that, we need to explore a minimum of the contrast parameter space: take data through both the round and bar occulters. To avoid spending too much time, priority is given to the long-wavelength (LW) channel as it will be the most strikingly better than the ground and popular for NIRCam. Hence, this activity does not make use of the short-wavelength (SW) channel.
The strategy is to use bright stars (K$\sim$5) to be very efficient and achieve high SNR and contrasts with reasonably fast readouts and in the least possible execution time, yet avoiding saturation limits and producing 10 to 100 frames. Having a large enough number of frames allows for frame selection (e.g remove the ones affected by Cosmic Rays and for aggressive post-processing with more degrees of freedom (KL). 

The way \coronsup was carried out is explained in greater details in Kammerer et al. 2022 (this conference)\cite{kammerer2022spie}: HD 114174 (with its WD companion, observable in June/July 2022) was set to be our \enquote{science scene}/star (Obs 1 \& 2 for MASK335R, Obs 5 \& 6 for MASKLWB with 2 rolls). 3 reference stars were carefully chosen (all main sequence G stars in the long baseline interferometry calibrator catalog to avoid multiplicity, all within K$\sim$5.0 and 5.2):  HD 111733, HD 115640, HD 116249 (all observed with a 9-point small grid dither "SGD"\cite{soummer2014_sgd} pattern) respectively at an angular distance of 5.3\degr, 4.6\degr and 11.9\degr from HD 114174 which would allow to explore difference time and pitch angle (solar elongation) baselines. At this point, we have not completed all this analysis (to come in a subsequent paper) because we focused on the readiness of the mode and thus mainly on the MASK335R/F335M setup.

\begin{figure}[h!]
\begin{center}
\includegraphics[width=0.95\textwidth, angle=0]{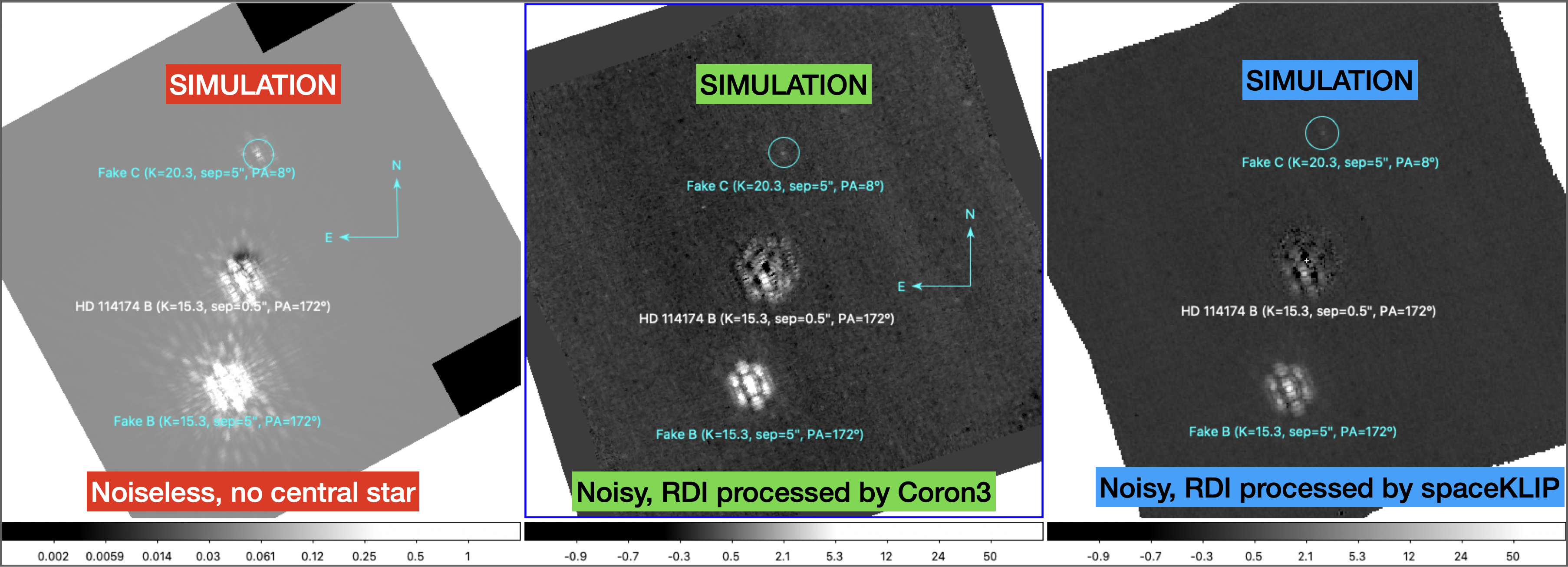}
\vspace{0.25cm}
\caption{Simulations of the HD 114174 scene performed with \pynrc and \webbpsfext. HD 114174 B was added with a separation of 0.5\arcsec and a position angle of 172$\degr$. In addition we added two other fake objects: \enquote{Fake C} to the north, 100 times (5 magnitudes) fainter to test our ability to recover point sources in the background limited regime and \enquote{Fake B} which is to the south at 5\arcsec separation which is a copy of  HD 114174 B, a probe to test our ability to recover its photometry close in versus far out. Left: the noiseless (slope image) scene without central star to show how HD 114174 B is affected by the coronagraph throughput unlike \enquote{Fake B}. Center: the noisy simulation RDI processed by the official \texttt{coron3} pipeline. Right: the same noisy simulation RDI processed by the custom \spaceklip community pipeline.}
\label{fig:wdsim}
\end{center}
\end{figure}

\section{Commissioning NIRCam Coronagraphy}
\label{sec:com}  % \label{} allows reference to this section

If we know the positions of all focal plane masks, manage to center the star in the TA aperture (after a decent telescope initial pointing), then the placement accuracy of the star behind the masks only depends on the small angle maneuver (SMA), the last offset of a few arc seconds. Unfortunately it is not exactly as simple as that. The calculation of the SMA is affected by the residual distortion and the positions of the masks have a significant uncertainty $\sim$5-10 mas (poor flat field quality, filter shifts, convolution by the PSF). Finally the centering accuracy in the TA aperture if severely affected by the measurement uncertainty of the centro\"iding algorithm which is more biased along the horizontal axis (x) because of the geometry of the PSF (wider than tall). 

\subsection{Astrometry and Distortion}
\label{sec:dist}  % \label{} allows reference to this section

\begin{figure}[h!]
\begin{center}
\includegraphics[width=0.8\textwidth, angle=0]{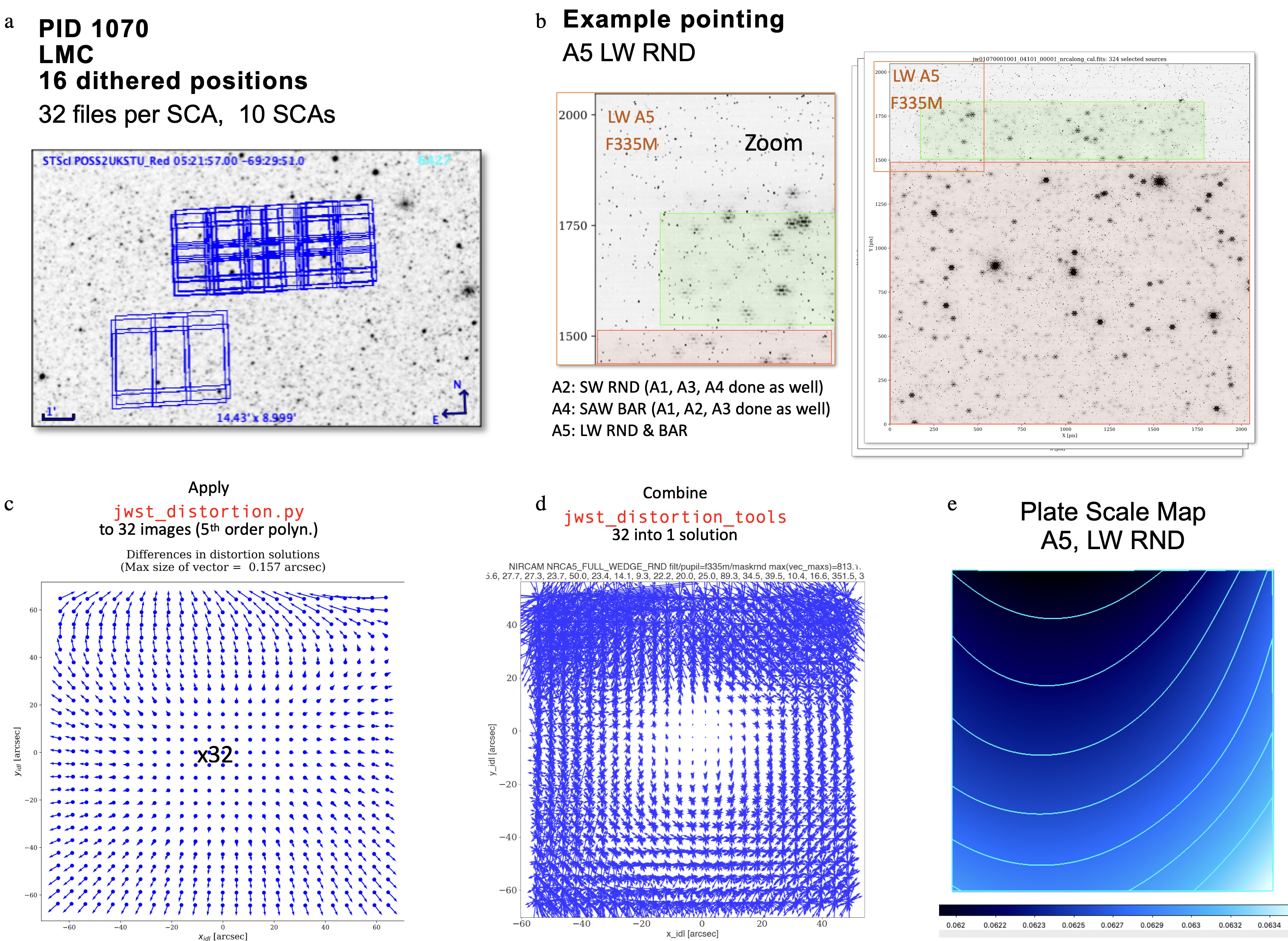}
\vspace{0.25cm}
\caption{a: paw prints of the 16 dithers of \coronlmc with NIRCam and FGS on a region of the LMC for which a precise Hawk-I/HST catalog is available\cite{sahlmann2019_catalog}. b: an example of A5 LW RND pointing. The discontinuity between the COM area and the rest of the full frame Coronagraphic field of view. c: we measure distortion using as many matching stars as possible using the \jwstdistortion suite with a 5$^{th}$ order polynomial fit. d: a higher quality solution is obtained combining all 32 measurements with a dedicated routine developed during commissioning for the Imaging Astrometry. e: plate scale map obtained for the A5 LW RND setup.}
\label{fig:astrom1}
\end{center}
\end{figure}

Figure~\ref{fig:astrom1} shows our procedure to analyse astrometric data (\coronlmc) on the Large Magellanic Cloud (LMC)\footnote{Custom routine to combine all distortion measurements: {\small \tt \href{https://github.com/arminrest/jwst_distortions_tools}{github.com/arminrest/jwst\_distortions\_tools}}}.

\begin{figure}[h!]
\begin{center}
\includegraphics[width=0.85\textwidth, angle=0]{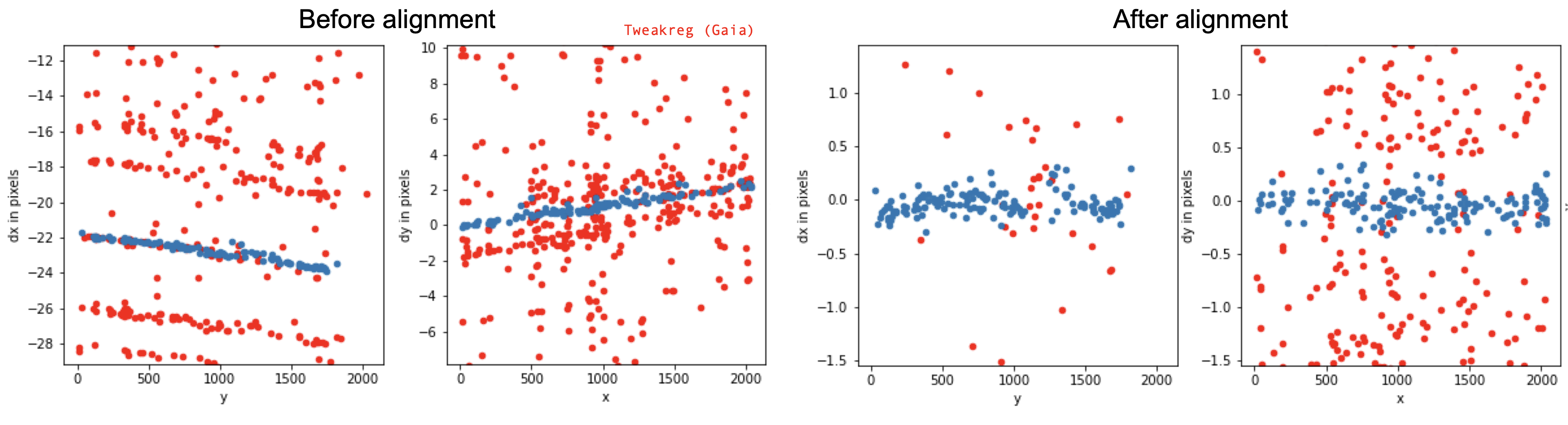}
\vspace{0.25cm}
\caption{Left (2 plots): Gaia matching using \texttt{tweakreg} in x and y. The blue dots are the stars that matched. There are significant shifts (especially in x, seen here has a $\sim$ 22-pixel vertical offset) and a rotation term (non-zero slope). Right (2 plots): after alignment, the selected blue points show a vertical spread of 3 to 4 mas in the region outside the COM (y$\leq$ 1100). The region around the discontinuity shows a greater spread. The distortion residuals well into the COM (y$\geq$ 1500) are sub 5 mas as well.}
\label{fig:astrom2}
\end{center}
\end{figure}

Subsequently, members of our team managed to adapt the DrizzlePac's TweakReg module\footnote{\texttt{tweakreg}: {\small \tt \href{https://drizzlepac.readthedocs.io/en/latest/_modules/drizzlepac/tweakreg.html}{drizzlepac.readthedocs.io/en/latest/\_modules/drizzlepac/tweakreg.html}}} to align our images taken at a given epoch to Gaia DR3 (taken as the on-sky truth). Figure~\ref{fig:astrom2} shows the improvement before (left) and after this alignment. This is also a great assessment of our distortion correction residuals (spread of the blue points) which are of the order of 5 to 8 mas RMS in the COM area (close to the coronagraphic masks) and about 3 to 4 mas in the rest of the full frame SCA (A5 or any other).  This means \nircam Coronagraphy can already be used for astrometric followups of point sources, orbital fitting and the  determination of model-independent dynamical masses of planetary-mass companions in synergy with other high contrast instruments. Nevertheless we hope in the future to improve these distortion correction residuals to 3 to 4 mas in the COM area as well. This will require to model out the discontinuity which is seen in the data (not many matches between y=1100 and y=1400, third plot of figure~\ref{fig:astrom2}).

\subsection{Flat fields and mask positions}
\label{sec:masks}  % \label{} allows reference to this section

We do not see the focal plane masks (even with a bright star is close) unless we can \enquote{back illuminate} them. We tried to median combine all dithered position of the LMC astrometric field to identify the mask positions but that turned out to be sub-optimal. On-sky flats taken with the zodiacal light allowed us to get a coarse measurement of the on-sky positions of all masks (and the COM \enquote{real estate}: ND squares, etc.). The issue is that flats were taken using wide bandpass filters (to collect enough light) and we had to measure filter to filter shifts and take them into account in the final mask positions, adding to the uncertainty.  

\begin{figure}[h!]
\begin{center}
\includegraphics[width=0.7\textwidth, angle=0]{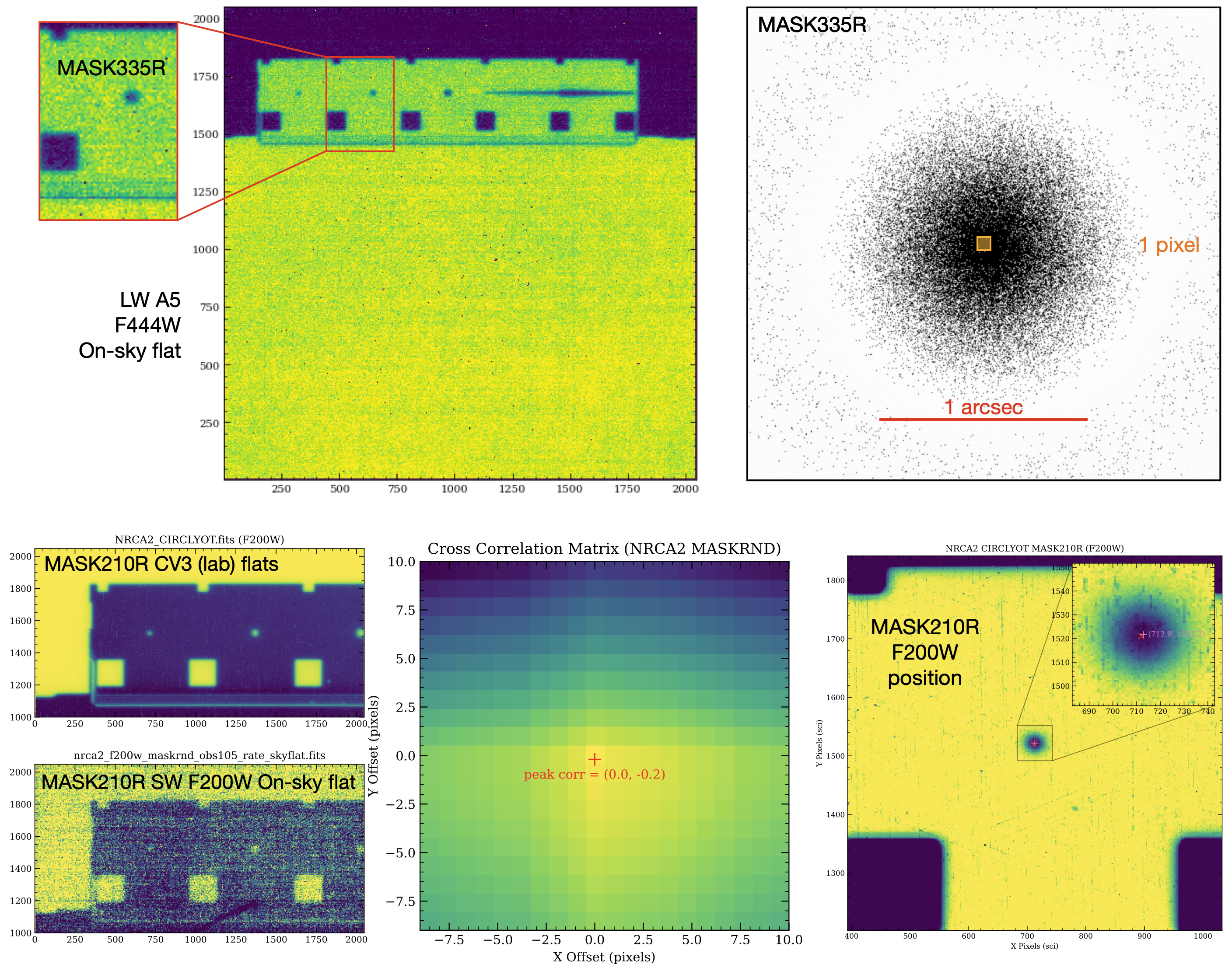}
\vspace{0.25cm}
\caption{Top left: the best (second attempt) on-sky coronagraphic flats taken during commissioning using the zodiacal light and the F444W filter. The COM \enquote{real estate} looks fuzzy because it is convolved with a somewhat large PSF FWHM. Top right: the size of a LW A5 pixel with respect to the MASK335R we are trying to locate. Bottom: a cross-correlation between ground (CV3) flats with high SNR and the flight flats using the zodiacal light allowed us to have a a rough estimate to $\sim$0.2 pixel of their position in detector coordinates. Naturally, the COM features looks sharper in the SW than in the LW.}.
\label{fig:flats}
\end{center}
\end{figure}

\begin{figure}[h!]
\begin{center}
\includegraphics[width=0.95\textwidth, angle=0]{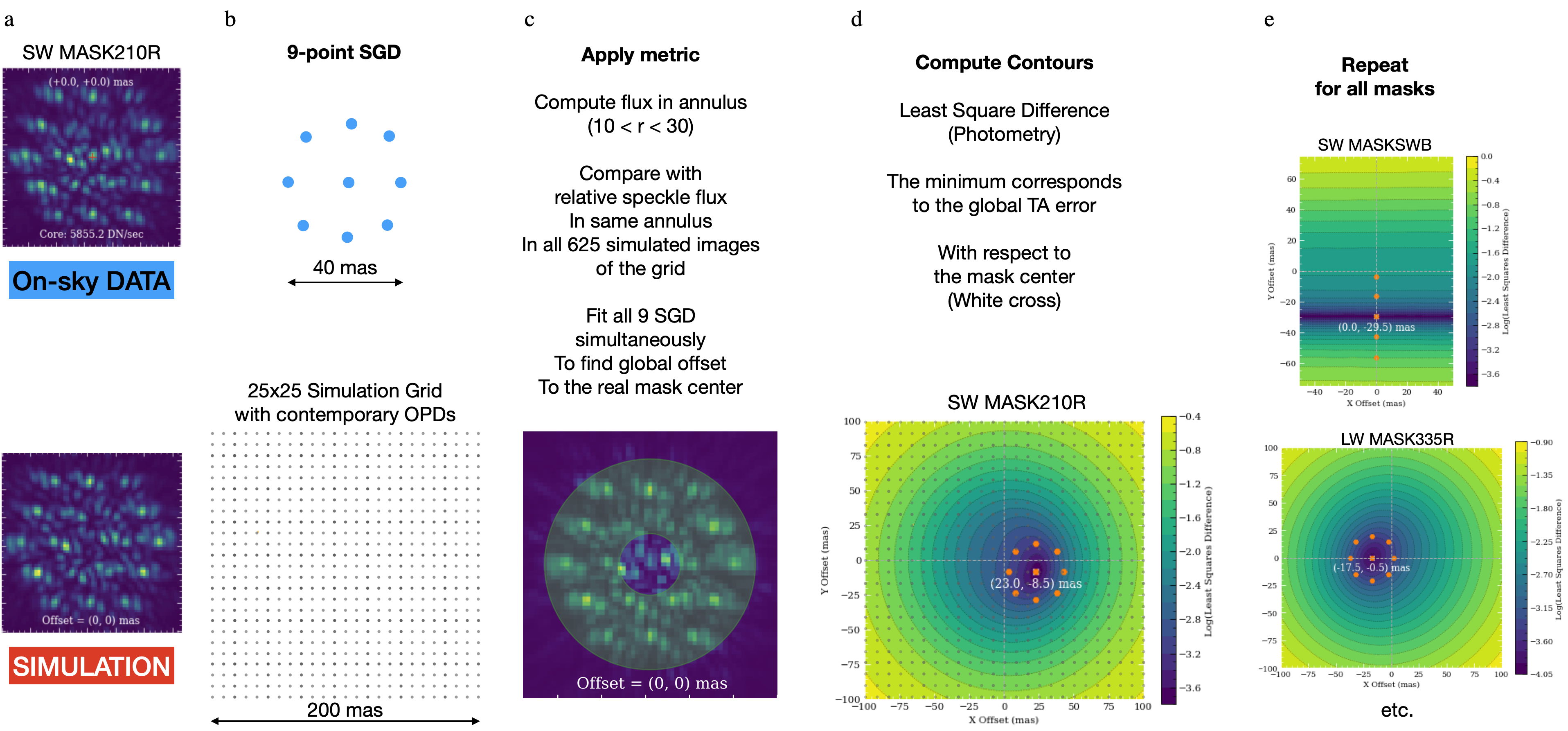}
\vspace{0.25cm}
\caption{Procedure to find each real mask center fitting all observations for a given TA (here, the 9-point SGD images of the MASK210R \coronta) to a grid of 625 simulated images (\pynrc and \webbpsf using contemporary OPDs). The metric we converged to is photometric, the relative flux of speckles in an annulus (c). Using least square difference and contours (d) we can determine how far from the mask center the 9 SGD images were and thus, the TA error. We then repeat this procedure for other masks. For the bar masks, we only had the 5-point vertical SGD, therefore only being able to access the y-axis offset, which is the one that matters.}
\label{fig:maskpositions}
\end{center}
\end{figure}

\subsection{NIRCam Coronagraphic Target Acquisition}
\label{sec:ta}  % \label{} allows reference to this section

\begin{figure}[h!]
\begin{center}
\includegraphics[width=0.95\textwidth, angle=0]{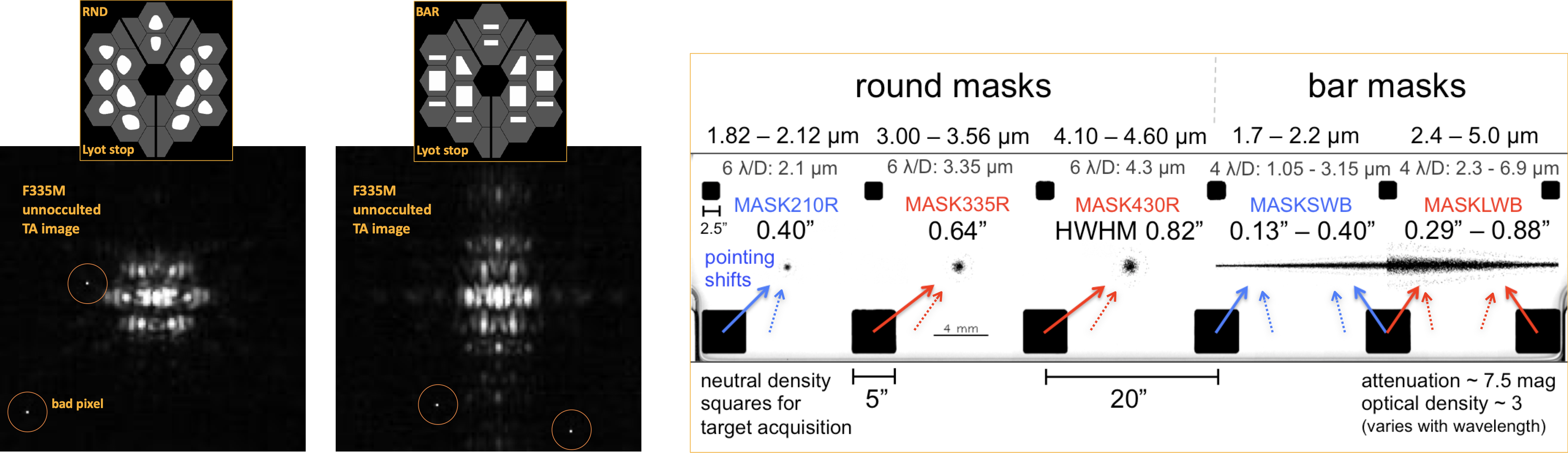}
\vspace{0.25cm}
\caption{Left: Lyot stop shapes for round and bar masks and resulting PSF (on-sky unnocculted TA images) through the F335M filter. Right: TA principle and SAM for each mask/case. For bright stars (typically K$\leqslant$6.3 mag, the TA centro\"iding is done through ND filters (black squares) and the SAM is represented by a plain arrow (blue for SW and red for LW). For fainter star TA, centro\"iding is done on the side of each ND and the SAM is represented by a dotted arrow.}
\label{fig:ta}
\end{center}
\end{figure}

\begin{figure}[h!]
\begin{center}
\includegraphics[width=1.0\textwidth, angle=0]{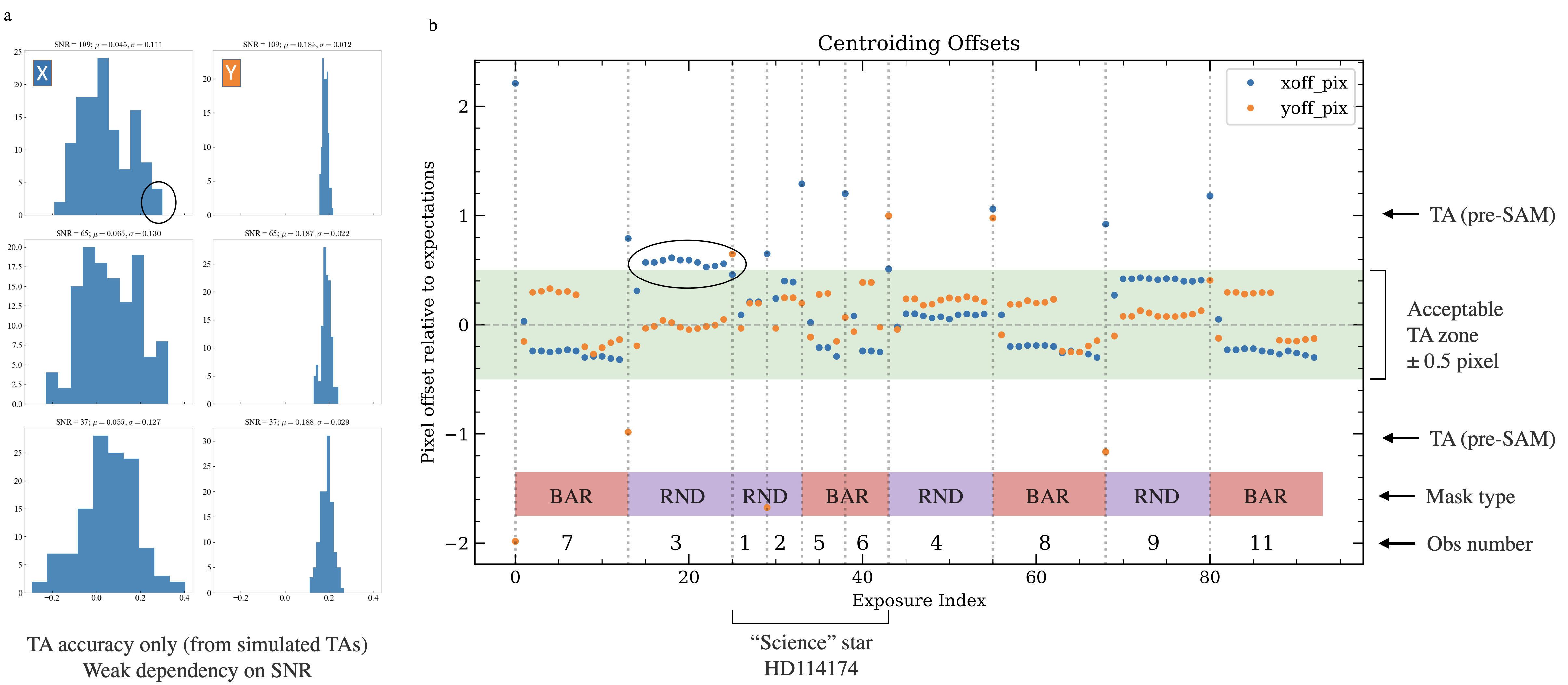}
\vspace{0.25cm}
\caption{a: simulated TA centro\"iding error distribution at three different SNR for the MASK335R and using a clone of the flight algorithm. It is clear that the current algorithm performs more poorly in x than in y. b: Analysis of the TA and all star and small grid dither positions for the program \coronsup (only LW) with respect to where we now think the masks centers are. Only Obs 3 is slightly out of the $\pm$0.5 pixel tolerance, likely in the tail of the distribution on the top left (high SNR, high centro\"id error). It corresponds to the blue curves in figure~\ref{fig:contrast-335r}.}
\label{fig:ta1441}
\end{center}
\end{figure}

The goal of coronagraphic target acquisition (TA) with NIRCam is to accurately align an astronomical point source—the \enquote{host}—on a coronagraphic mask (occulter). Coronagraphic TA involves an initial slew of the telescope to place the target on a 4\arcsec$\times$4\arcsec subarray in the ~4\arcsec vicinity of the selected mask. If the target is bright than (K$\leqslant$6.3), the subarray is located behind a neutral density square (nominally ND $\sim$ 3). If fainter (K$\geqslant$6.3), the target is positioned behind a nearby, clear (ND = 0) region of the coronagraphic optical mount (COM). The first phase of TA is complete when the detector obtains an exposure of the target on an appropriate region of the COM (ND = 0 or 3) near the specified coronagraphic mask. Coronagraphic TA images are always be taken in either the F210M or F335M filter, for short- or long-wavelength (SW, LW) coronagraphy, respectively. Coronagraphic TA images are taken using 128$^2$ or 64$^2$ subarrays, for SW or LW, respectively. Figure~\ref{fig:ta} shows the principle of the Coronagraphic TA with each mask's position and inner-working-angle (IWA). In figure~\ref{fig:ta1441} we see that the shape of the unnocculted PSF affect the accuracy and repeatability of the TA with the current centro\"iding algorithm. In section~\ref{sec:rec} we discuss the perspectives for improvements.

The telescope pointing accuracy is superb. The initial slew always placed our star within $\sim$4 pixels of the TA subbarrays (depending on the \enquote{quality} of the guide star used by the Fine Guiding Sensor\cite{doyon2012_FGS}, FGS 1 or 2. Some guide stars have Gaia\cite{gaia2021_dr3_long} information and other are not). Globally performance indicators related to TA that we were able to assess are:
\bi
\item Target Positioning – TA Performance: TA on the coronagraphic PSF can induce up to 20 mas offsets from true center, which propagates through an observation’s pointing. Likely due to PSF side lobes (figure~\ref{fig:ta}) interfering with center of mass algorithm. Offsets from TA to occulter are consistent with a sigma of $\sim$ 3 mas.
\item SAM performance: Even after accounting for TA inaccuracies, there are residual errors in the source positioning relative to the commanded offset position. For both masks, the star misses the expected location in a consistent manner (the FGS moved the telescope to the same location every time). The post TA stellar positionning would have consistently missed the specified reference mask position by less that 0.5 pixels (by $\sim$30 mas) in x and/or y. Worse performance at MASKLWB locations (for each filter central wavelength), which is closer to the corner of the \nircam field of view, suggesting distortion corrections could be the culprit, likely due to differences induced from pupil wheel tuning activities.
\item SGD Performance: we measured SGD offsets by performing a cross-correlation of the central SGD position with all surrounding dithers as well as performing cross-correlations with the \webbpsf simulation of a perfectly centered source. SGD perform very close to expectations, approximately within $\sim$2 to 3 mas from the ideal locations. These measurements are consistent with what the FGS team has been reporting.
\ei

\subsection{LW Pupil Wheel Alignment, 1.5 million km from Earth!}
\label{sec:pw} 

Half way through our TA commissioning activities, we noticed all LW coronagraphic (occulted) images had a unexpected pattern. After discarding the occasional red companion (because the bright speckle on Figure~\ref{fig:pw} did not show for SW), we ran extensive \pynrc simulations with various amounts of pupil shear. These simulations revealed that the LW pupil wheel (PW) rotation needed to be adjusted.  We triggered an anomaly (standard procedure in the event of a serious issue) to make sure we would get the necessary ressources and assistance from OSS (Operations Scripts System). Indeed, the software was not design to perform such an alignment procedure on orbit (07C as \enquote{contingency} in figure~\ref{fig:nrc-cars}). We had to think of several strategies and discuss them extensively. We finally had to proceed with a safe two-step approach:

\begin{figure}[h!]
\begin{center}
%\includegraphics[width=0.45\textwidth, angle=0]{FiguresSPIE/pseudo_SGD_contrast_Obs15_MASK210R.png}
%\hfill
\includegraphics[width=0.95\textwidth, angle=0]{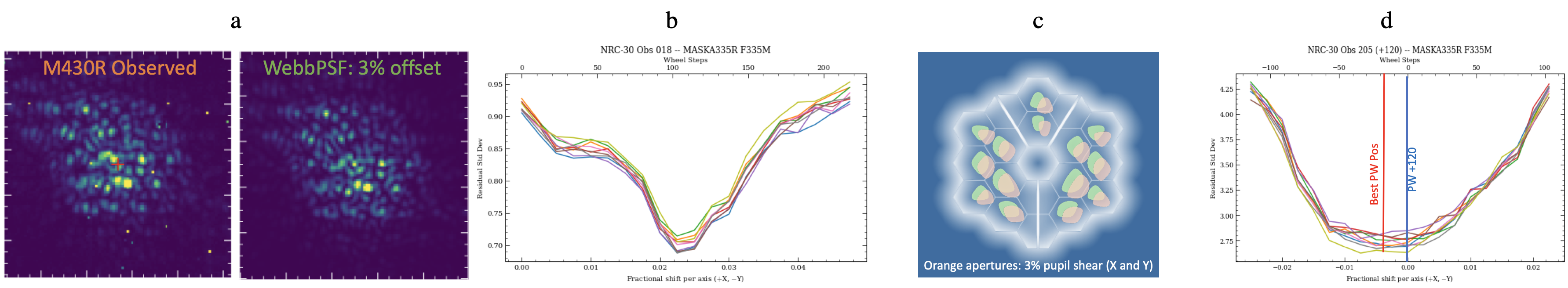}
\vspace{0.25cm}
\caption{Pupil shear (misalignment of the round Lyot stop in this case). a: Observed (left) and simulated (right) occulted coronagraphic PSF (MASK430R but MASK335R was similar); b: analysis of all SGD position and assessment of the fractional shift (in \%) of the pupil. The ideal position is 0.00. c: visualization of a 3\% pupil shear (in both X and Y) with respect to the diffracted light spatial distribution in the pupil plane. d: analysis of the fractional shift (all SGD) after a PW rotation of +120 steps, close to optimal and the trade-off position we decided to keep to accommodate both round masks (known offset).}
\label{fig:pw}
\end{center}
\end{figure}

\begin{enumerate}
\item Measure the field offsets ($\sim$7 pixels expected) caused by the PW rotation at 6 discrete positions, with coronagraphic optics inserted but unnocculted as the Engineering Imaging Template (only one allowing to perform table loads to rotate the PW) did not have any TA.
\item Load the measured offsets (transposed in ideal coordinates) into a subsequent observation with Coronagraphic TA for the MASK335R and MASKLWB cases and compare the PSF pattern with simulations done using contemporaneous OPD maps.
\end{enumerate}

Unfortunately a significant tilt event occurred on June 27 2022 just when we were finally able to acquire data in occultation. Using contemporaneous OPD maps to perform our analysis against the most realistic simulations was intense (short time and high pressure) but primordial. We converged on two trade-off LW PW offsets: +120 steps for the RND masks and +105 steps for the BAR mask.

\begin{figure}[h!]
\begin{center}
\includegraphics[width=0.85\textwidth, angle=0]{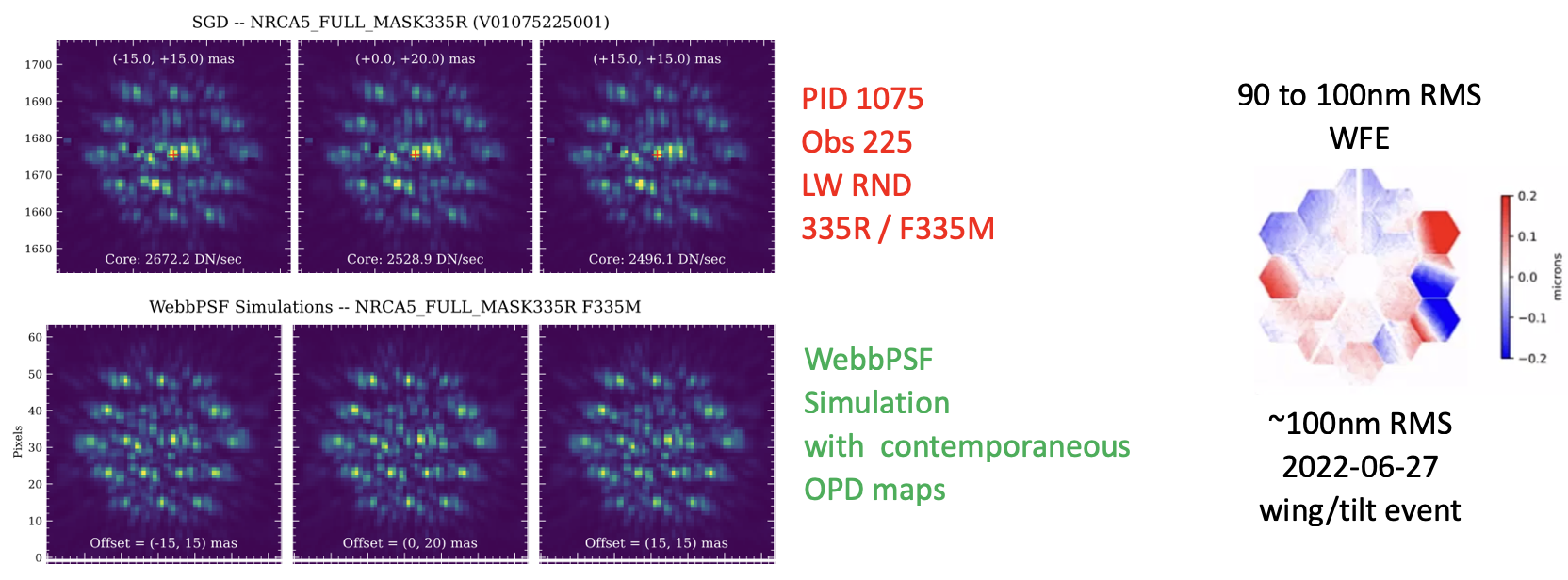}
\vspace{0.25cm}
\caption{Tilt event of the June 27 2022 causing the global RMS wavefront error to increase to $\sim$90-100 nm (right). This event occurred while we were attempting to align the LW PW and assess the mask positions.}
\label{fig:tilt}
\end{center}
\end{figure}

Based on these observations we could move on to \coronsup. We knew the TA would probably have degraded slightly with all the successive coordinate transformations and shift measurements and the absence of new astrometric/distortion data since we had moved the two LW pupil wedges and hence the associated distortion solutions by a small amount. 

\section{Contrast Performance}
\label{sec:coron-perf} 

Figure ~\ref{fig:on-sky-335r} showed the ability of the mode to recover cleanly a faint companion inside the so called inner-working-angle (IWA) of the coronagraph. In this section we will focus on the performance of the LW round mask MASK335R. Performance of the LW bar mask MASKLWB are good as well and reported in Kammerer et al. 2022 (this conference)\cite{kammerer2022spie}.

\subsection{Data reduction and post-processing}
\label{sec:data}  % \label{} allows reference to this section

All the results presented in this paper are using the \jwst pipeline that will be described in Gordon et al. 2022b (in prep.)\cite{gordon2022_JWSTPIPE} stage 1 (\texttt{detector1}) and 2 (\texttt{image2}) unless specified otherwise. Following the recommended strategy and implementation of the coronagraphy-specific (\nircam and \miri) stage 3 (\texttt{coron3}), a mini-PSF reference library is build using the small grid dither (SGD)\cite{soummer2014_sgd} of one or several reference stars. A principal component analysis approached is then used, the so-called he Karhunen-Lo\`eve Image Projection (KLIP)\cite{soummer2012_klip} to subtract an optimal reference PSF to the science scene and reveal faint signal around it. \spaceklip is an agile community version of \texttt{coron3} based on the very successful and popular \pyklip. It is in active development\cite{kammerer2022spie} and has many functionalities that we are not describing in this work: forward modeling of point sources and disks, simultaneous astrometry and photometry using a Monte-Carlo approach, etc.

Here we present the performance to the best of our knowledge at the current state of our analysis just after the science readiness review with a limited analysis period after the \enquote{Coronagraphic Suppression Verification} data (\coronsup) was taken early July. These analysis focused on the MASK335R as it was the setup chosen to meet the readiness criteria. We anticipate that performance will evolve favorably as TA errors will be reduced (during Cycle 1) and TA repeatability can be improved, improving the TA centro\"iding algorithm (no timeline for now).

\begin{figure}[h!]
\begin{center}
%\includegraphics[width=0.45\textwidth, angle=0]{FiguresSPIE/pseudo_SGD_contrast_Obs15_MASK210R.png}
%\hfill
\includegraphics[width=0.9\textwidth, angle=0]{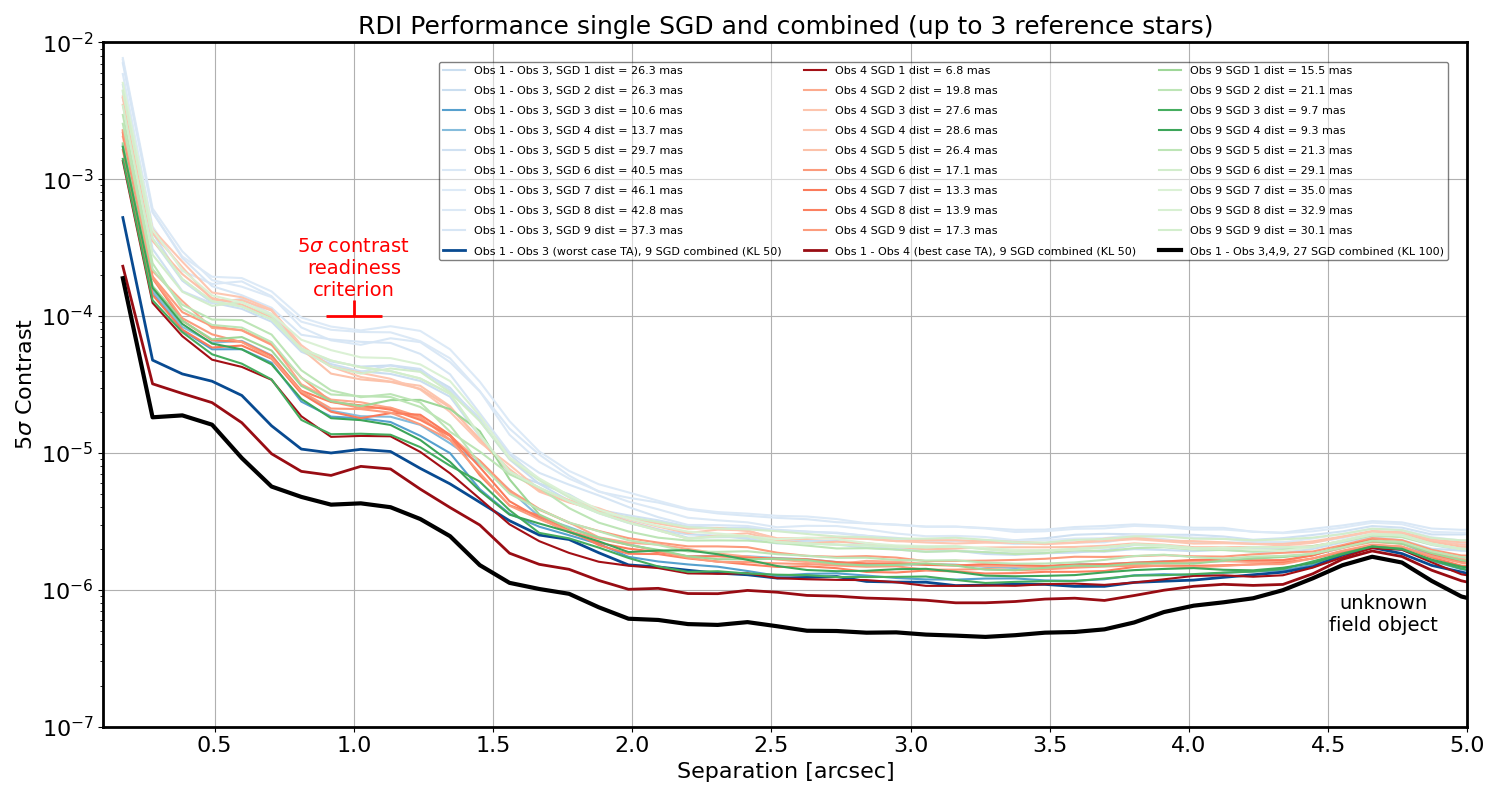}
\caption{RDI contrast using \coronsup Obs 1 (HD 114174) as \enquote{science scene} and individual small grid dither as reference (thin color lines). The closer (in TA error) the reference the better the contrast. The thicker blue line is the contrast obtained combining the 9-point SGD of Obs 3 (worst case). The thicker blue line is the contrast obtained combining the 9-point SGD of Obs 4 (best case). The thick black line is the contrast obtained combining all 27 SGD (3 reference stars) in a single PSF library with KLIP and using 100 principal components. Note: these azimuthal contrast curves are affected by the presence of HD 114174 B at 0.5\arcsec and by the unknown field object at 4.7\arcsec which have not been masked out. The ND throughput has been corrected but the coronagraphic mask throughput has not.}
\label{fig:contrast-335r}
\end{center}
\end{figure}

In the {\sl Characterization of JWST science performance from commissioning} report by Rigby et al. 2022\cite{rigby2022long} posted on July 12 2022\footnote{A report on the actual JWST science performance, as characterized through the 6-month commissioning activities: {\small \tt \href{https://jwst-docs.stsci.edu/breaking-news}{jwst-docs.stsci.edu/breaking-news\#BreakingNews-JWSTscienceperformancereportattheendofcommissioning}}}, the same contrast curves are shown as in figure~\ref{fig:contrast-335r} but only displaying the 9 SGD of the worst case reference star (Obs 3). Here we also applied a corrective factor of $\sim$2 to take into account the wavelength dependence of the TA neutral density (ND) filter \footnote{The transmission curve of the ND is shown in Kammerer et al. 2022 (this conference)}. In the end, the contrast in this paper is thus about twice worst than in the report. Figure ~\ref{fig:adirdi} shows corrected contrasts and compares with that of the commissioning report.

\subsection{Long Wavelength performance}
\label{sec:lw}  % \label{} allows reference to this section

The LW achievable contrast (MASK335R) is summarized in figures~\ref{fig:contrast-335r} \&~\ref{fig:adirdi}.

\subsection{Short Wavelength performance}
\label{sec:sw}  % \label{} allows reference to this section

\begin{figure}[h!]
\begin{center}
\includegraphics[width=0.85\textwidth, angle=0]{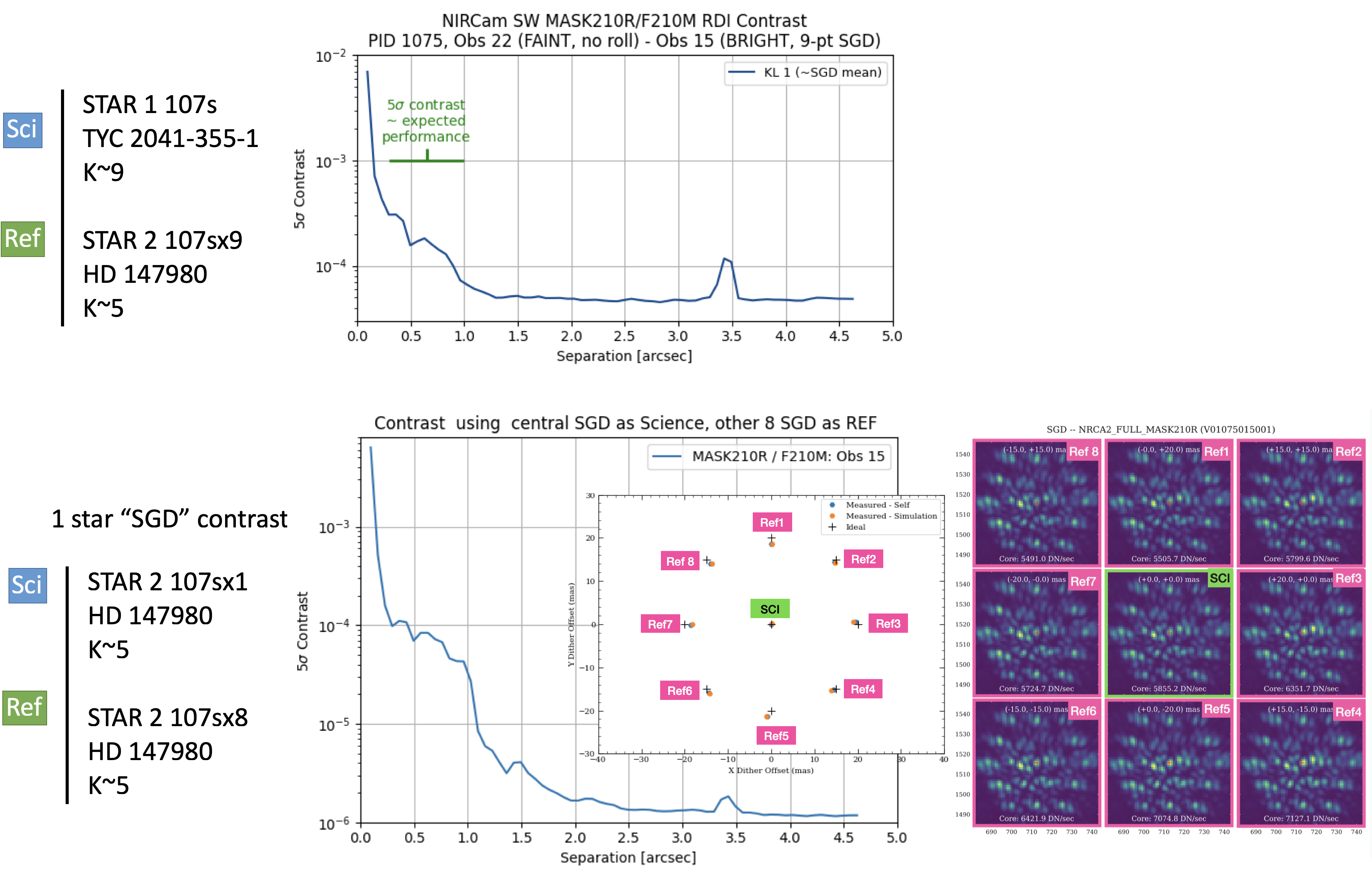}
\caption{Top: Preliminary MASK210R contrast obtained during the TA commissioning activity (\coronta) using a single 10-second exposure (Obs 22) on a rather faint star (K$\sim$9) as \enquote{science scene} and a brighter star (K$\sim$5, Obs 15) with 9-point SGD (100s each) as a PSF reference for KLIP, displaying only the curve obtained with 1 principal component (KL 1).   Bottom: Small Grid Dither \enquote{Self Reference Contrast} for the MASK210R/F210M setup (\coronta Obs 15). Here we are in the ideal case (no OPD difference, only TA errors) as we subtract to the central position the mini library of the other 8 SGD positions.}
\label{fig:SGDcontrast}
\end{center}
\end{figure}

\coronsup (Suppression Verification) did not include any SW measurements so we only had a few images from \coronta (Coronagraphic TA) to experiment and compute less-than-ideal preliminary, yet encouraging contrasts. Figure ~\ref{fig:SGDcontrast} shows the results of these experiments. We also subtracted a synthetic PSF library to each SGD to see which one would be closest to the ideal positioning. It revealed to work but was judged as a marginally sensitive approach (only sensitive between 0.2\arcsec and 0.55\arcsec) for our main goal at the time: assessing the relative positioning with respect to the mask.

\section{Discussion}
\label{sec:discussion} 

%\subsection{Commissioning hurdles}
%\label{sec:hurdles}  % \label{} allows reference to this section

\subsection{PSF Subtraction Strategies}
\label{sec:adirdi}  % \label{} allows reference to this section

\begin{figure}[h!]
\begin{center}
\includegraphics[width=0.8\textwidth, angle=0]{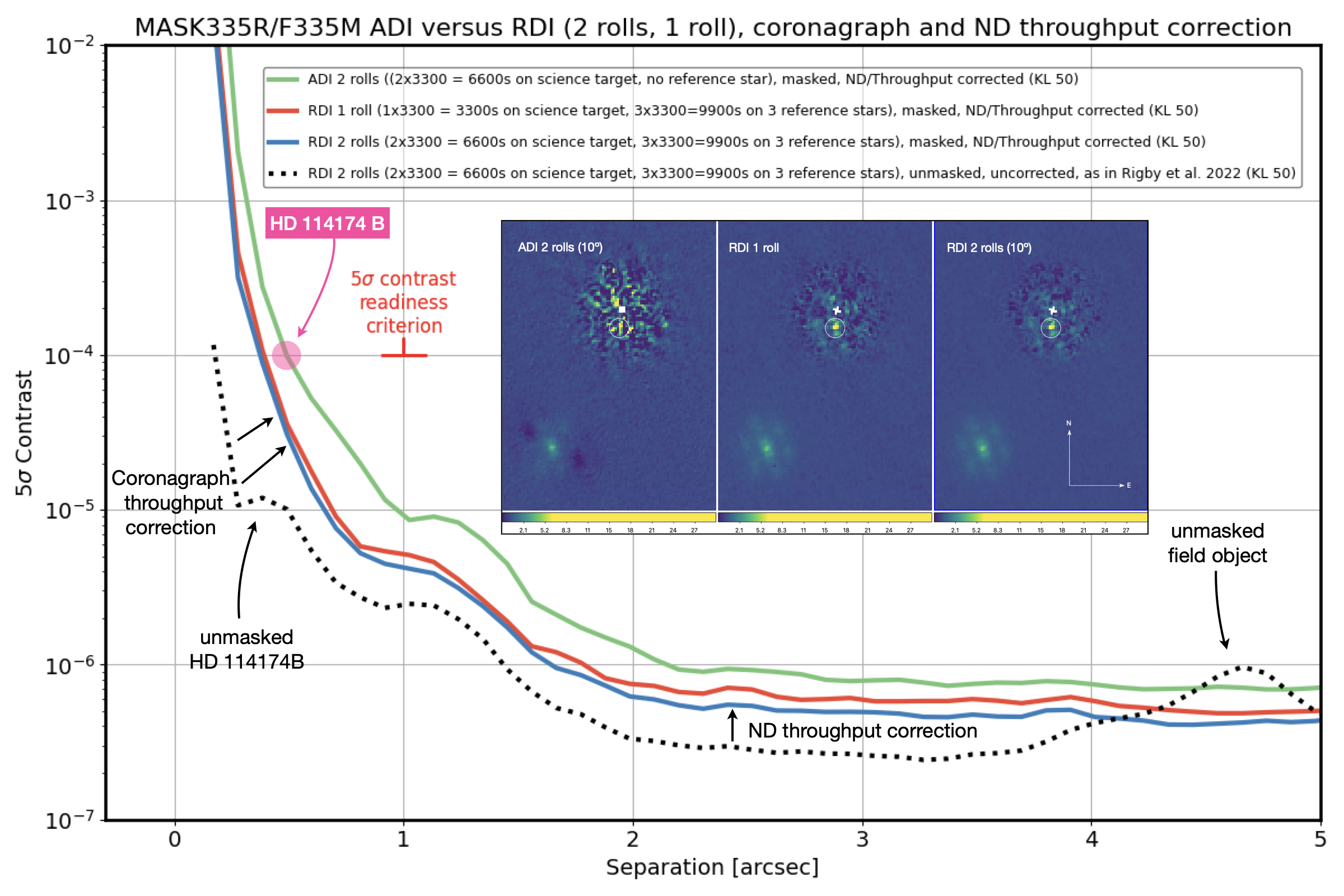}
\caption{ADI (Roll subtraction with a 10\degr roll angle) yields good results beyond 1\arcsec, very good results beyond 2\arcsec. RDI without roll subtraction but with one or several reference stars taken with SGD yields excellent results even within the IWA. A second roll with a 10\degr offset only marginally improves the sensitivity, mainly outside the IWA (0.63\arcsec here). Here we compare our new MASK335R F335M RDI curve with that of Rigby et al. 2022\cite{rigby2022long} (Figure 12) that was not corrected for ND and coronagraph throughputs and for which the two detected point sources were not masked when azimuthally computing the contrast.}
\label{fig:adirdi}
\end{center}
\end{figure}

For Cycle 1 and probably Cycle 2, \textbf{we recommended users to adopt as main coronagraphic strategy: 2 rolls (as separated as possible in the limit of 14\degr) and at least 1 PSF reference star taken back to back in an uninterruptible sequence}. Our \coronsup commissioning program allowed us to start investigating what the best and most efficient PSF subtraction strategy would be with \nircam coronagraphs given the telescope stability, wavefront residuals and what we know of our TA error and repeatability issues.  Figure ~\ref{fig:contrast-335r} shows the importance of the SGD mitigation strategy: the more reference star and diversity in positioning the merrier. In general equating the science target SNR with a standard star with a 9-point SGD pattern allows a very clean detection at 0.5\arcsec (4.7 $\lambda/D$, well within the 6 $\lambda/D$ IWA) of a $10^4$ contrast point source companion. Figure ~\ref{fig:adirdi} shows that having one or several reference stars is more important than performing rolls if the goal is to detect a faint companion or structures as close as possible (e.g. between 0.4\arcsec and 1\arcsec in the speckle limited regime). 

\subsection{Operational Maturity}
\label{sec:ops-maturity} 

Even though we identified areas to improve (mainly the TA but also the astrometric analysis) the operational maturity of the mode is high. Indeed we have performed many TAs and reached a point where guide star acquisition (from FGS) only fails once in a while, not more than for other JWST modes. 

With new \nircam reference files (e.g. distortion reference files for Coronagraphy) have recently (July 2022) been delivered to the Calibration Reference Data System (CRDS)\footnote{JWST Calibration Reference Data System (CRDS): {\small \tt \href{https://jwst-crds.stsci.edu}{jwst-crds.stsci.edu}}}, the official \jwst pipeline stage 3  (\texttt{coron3}) is now able to provide \texttt{i2d.fits} in MAST which are more than a quicklook and thus practically \enquote{science grade}, very similar to the \spaceklip images that we have shown in this paper. Intermediary products are fantastic: For instance, the \texttt{psfsub} product is a cube for each science integration (53s for \coronsup) as a slice for each roll, KLIP-subracted (RDI) by one of the reference stars. The WD companion HD 114174 B is clearly and cleanly imaged in nearly every frame.

The Exposure Time Calculator (ETC)\footnote{JWST Exposure Time Calculator (ETC): {\small \tt \href{https://jwst.etc.stsci.edu}{jwst.etc.stsci.edu}}} predicted the percetage of the full well with less than 20\% discrepancy with the reality. The TA SNR were as expected. ETC is overoptimistic (no TA error, less detector noise) but the current ETC OPD is worse and there is no SGD implementation. All in all, the ETC is currently still adequate to prepare proposals and observations. 

JDox needs to be updated as for all the JWST mode. \nircam Coronagraphy presented outdated contrast curve predictions\cite{beichman2010}. Nevertheless the recommended PSF recommended strategy and the HCI articles still hold.

\subsection{Commissioning hurdles, recommendations and perspectives}
\label{sec:rec}  % \label{} allows reference to this section

The \nircam focal plane masks are forgiving but it is challenging to know where they are with respect to our star of interest. 
The main hurdle has been the LW pupil wheel misalignment which made us slip by about 4-5 weeks. Astrometry with the Coronagraphic PSF and 10 SCAs was a challenge but the team now has the software infrastructure and knowledge to do it again. Our simulation framework (\pynrc, \webbpsf, \webbpsfext) has allowed us to come up with agile ways to analyze data and move one with the commissioning of \nircam Coronagraphy to the level we are at now, offering the mode to Cycle 1 users, including \ers and \gto. \textbf{The Target Acquisition (TA) can be further improved. We recommend that the TA parameters be remeasured carefully during Cycle 1 which could afford significantly better performance than is currently possible and very much better than pre-launch predictions} (subject to the state of the OTE {\sl vis-\`a-vis} tilt events, etc.).

Our plan as of July 2022 is:
\begin{enumerate}
\item Proceed with the July observations (ERS etc.) and get even more knowledge on the TA repeatability
\item Plan a LMC astrometric calibration program similar to \texttt{NRC-21b} \coronlmc: we now have the tools to analyze this more quickly and determine the distortion solutions and offsets for both SIAF and CRDS updates, take into account the rotation term (possibly $\sim$0.1 deg from our estimations).
\item Perform a follow-up calibration program similar to \texttt{NRC-30} \coronta that checks TA accuracy is improved.
However to investigate robustness and repeatability we need more than 1 observation per mask and therefore, if it’s acceptable (likely better than now), we will not stop science programs after this.
\end{enumerate}

For future cycles we hope to be able to implement and support \nircam Coronagraphy with simultaneous SW and LW (always saving both, as for the Imaging mode). We think we will improve the official \jwst pipeline and \mast products based on all feedback we would have receive from the vibrant \hci community both in terms of pre-processing and post-processing. We wish to improve the TA algorithm which is currently a limitation for robustness and accuracy. Finally we wish to improve and fine tune our astrometric calibration method by eventually taking into account (via a model) the discontinuity between the two areas (COM and non-COM).

The RDI with SGD strategy appears very good. While it can be time consuming: slew to the reference star(s), SGD, it has been shown that \nircam contrasts are only marginally affected by a spectral mismatch between science target and reference star(s) and that brighter reference star(s) can be observed with the same or higher SNR in less total time. Finally the telescope wavefront monitoring has shown excellent stability over hours (about 20 nm over a huge slew, larger than our recommended $\sim$5-10\degr$\,$ typically slew between targets of a same program. In other words, \textbf{it is probably better in most cases to use brighter, slightly farther reference star(s) with slightly different spectral types than chose a roll only strategy}. 

Once many reference stars will have been observed using many settings, it will be possible to use an archive reference star library in the RDI KLIP subtraction. That reference star library can be ideally composed of many stars, eventually with/without tilt events (anything that can happen during a science observation). We have started to experience with using a synthetic PSF library (only with SW MASK210R F210M as shown on figure~\ref{fig:SGDcontrast}. Increasing the size of the library and the information in it (include tilt events, TA errors, spectral and brightness diversity) should yield good results. We can even think of using a hybrid PSF library composed of both real stars and synthetic PSFs (created using contemporary OPDs and non-contemporary features such as tilt events / segment relaxation, accounting for micrometeoroid impacts, etc.).

Based on our experience with \coronsup, there is absolutely no doubt that \nircam Coronagraphy will soon deliver impactful science results as the measured flight performance is above expectations\cite{carter2021, hinkley2022simulations}. Known giant exoplanets will be characterized and studied further and new planets will be discovered. Unlike AO on the ground (even with lasers) \nircam Coronagraphy can be used on faint hosts: brown dwarfs, galaxies, etc. Possibilities are endless. Finally, if SW and LW can be saved simultaneously in the future as well as with the use of archive and/or synthetic or hybrid PSF libraries, then \nircam Coronagraphy can become a more efficient JWST mode and yield even more science return. 

\acknowledgments % equivalent to \section*{ACKNOWLEDGMENTS}       

 These observations were made possible through the efforts of the many hundreds of people composing the international commissioning staff of JWST. This work is based on observations made with the NASA/ESA/CSA James Webb Space Telescope. The data were obtained from the Mikulski Archive for Space Telescopes at the Space Telescope Science Institute, which is operated by the Association of Universities for Research in Astronomy, Inc., under NASA contract NAS 5-03127 for JWST. These observations are associated with programs \coronlmc, \coronta and \coronsup. Jens Kammerer is supported by programs PID 1194, 1411, and 1412 through a NASA grant from the Space Telescope Science Institute, which is operated by the Association of Universities for Research in Astronomy, Inc., under NASA contract NAS 5-03127. Some of the research described in this publication was carried out in part at the Jet Propulsion Laboratory, California Institute of Technology, under a contract with NASA. We thank warmly Johannes Sahlmann who was the main developer and initiator of \texttt{pystortion} precursor to \jwstdistortion and \pysiaf, making the first prototype of JWST astrometric calibration in NIRISS Imaging mode\cite{sahlmann2019}. We thank Scott Friedman, our Commissioning Scientist who kept us on a schedule and ran the fabulous JDB (JWST Daily Briefings). We also thank the SPIE Organizing Committee and the Proceedings Coordinators.
 %\footnote{See {\small \tt \href{https://jwst-docs.stsci.edu/display/JTI/NIRISS+Aperture+Masking+Interferometry}{https://jwst-docs.stsci.edu/display/JTI/NIRISS+Aperture+Masking+Interferometry}}}

\bibliography{SPIE2022} % bibliography data in report.bib
\bibliographystyle{spiebib} % makes bibtex use spiebib.bst

%\pagebreak

\vspace{-0.2cm}
%\section{Appendix}
\section{Acronyms}
%Acronyms used in this paper are listed in table~\ref{table:acronyms}. 
For more JWST related acronyms and abbreviations: {\small \tt \href{https://jwst-docs.stsci.edu/jwst-acronyms-and-abbreviations}{jwst-docs.stsci.edu/jwst-acronyms-and-abbreviations}}
%\footnote{For more JWST related acronyms and abbreviations: {\small \tt \href{https://jwst-docs.stsci.edu/jwst-acronyms-and-abbreviations}{jwst-docs.stsci.edu/jwst-acronyms-and-abbreviations}}}.
\vspace{-0.2cm}
\begin{table}[h!]
\small
%\caption{List of acronyms.}
%\vspace{-0.15cm}
\begin{center}
\begin{tabular}{|l|l|}
\hline
ADI & Angular Differential Imaging \\
ALMA & Atacama Large Millimeter/submillimeter Array \\
AO & Adaptive Optics \\
APT & Astronomer's Proposal Tool \\
COM & Coronagraphic Optical Mount \\
CRDS & (JWST) Calibration Reference Data System \\
CSA & Canadian Space Agency \\
CVT & Coronagraphic Visibility Tool \\
CV3 & Cryo-Vacuum test \#3 \\
DI &  Direct Imaging \\
DMS & Data Management System \\
ESA & European Space Agency \\
ETC & Exposure Time Calculator \\
FGS & Fine Guiding Sensor \\
FoV & Field of View \\
FPA & Focal Plane Array \\
Hawk-I & High Acuity Wide field K-band Imager (VLT)\\
HCI & High Contrast Imaging \\
HST & Hubble Space Telescope \\
JDox & JWST user documentation (shorthand) \\
JWST & James Webb Space Telescope \\
KLIP & Karhunen-Lo\`eve Image Projection \\
LMC & Large Magellanic Cloud \\
LW & Long Wavelength (Channel) \\
mas & milliarcsecond \\
MAST & Mikulski Archive for Space Telescopes \\
MIRI & Mid-Infrared Instrument \\
NASA &  National Aeronautics and Space Administration \\
ND & Neutral Density (filter) \\
NIRCam & Near InfraRed Camera \\
OPD & Optical Path Difference \\
OSS & Operations Scripts System \\
OTE & Optical Telescope Element \\
PSF & Point Spread Function \\
PI & Principal Investigator \\
PID & Proposal ID (Identification in APT and MAST) \\
PIL & Pupil Imaging Lens \\
PW & Pupil Wheel \\
RDI & Reference Differential Imaging \\
RMS & Root Mean Square \\
RND & Round (mask) \\
SAM & Small Angle Maneuver\\
SCA & Sensor Chip Assembly \\
SGD & Small Grid Dither(s) \\
SI & Science Instrument \\
SIAF & Science Instrument Aperture File \\
SNR & Signal to Noise Ratio \\
%STScI & Space Telescope Science Institute \\
SW & Short Wavelength (Channel) \\
TA & Target Acquisition \\
VLT & Very Large Telescope\\
WD & White Dwarf \\
WFE & WaveFront Error \\
YSO & Young Stellar Object \\
\hline
\end{tabular}
\end{center}
\label{table:acronyms}
\end{table}%

\end{document}